\documentclass[12pt]{article}

\usepackage{authblk}
\usepackage{helvet}
\usepackage[a4paper, total={6.5in, 10in}]{geometry}
\usepackage{footmisc}

\usepackage{graphicx}
\usepackage[usenames,dvipsnames,svgnames,table]{xcolor}

\def \L {\text{L}}
\def \M {\text{M}}
\def \HH{\text{H}}
\def \A {\text{A}}

\usepackage{amssymb}
\usepackage{amsfonts}
\usepackage{amsmath}
\usepackage[linesnumbered,ruled,vlined]{algorithm2e}
\usepackage{color}

\usepackage[font={small,it}]{caption}
\usepackage{subcaption}

\input{mysymbol.sty}

	\sffamily
	\title{\textbf{Functional Alignment with\\ Anatomical Networks is Associated with Cognitive Flexibility}} 
	
	\author[1]{John D. Medaglia}
	\author[2]{Weiyu Huang}
	\author[1]{Elisabeth A. Karuza}
	\author[1]{Sharon~L.~Thompson-Schill}
	\author[2]{Alejandro Ribeiro}
	\author[2,3]{Danielle S. Bassett}
	\affil[1]{Department of Psychology, University of Pennsylvania, Philadelphia, PA, 19104 USA}
	\affil[2]{Department of Electrical \& Systems Engineering, University of Pennsylvania, Philadelphia, PA, 19104 USA}
	\affil[3]{Department of Bioengineering, University of Pennsylvania, Philadelphia, PA, 19104 USA}
\date{}                     
\begin{document}
	
	\maketitle
	\newpage
	\section*{Article summary}
	Human cognitive flexibility emerges from complex dynamics in anatomical networks. However, no concise measure integrating anatomical and functional organization in support of cognitive flexibility has been identified. We examine cognitive ``switch costs" -- a measure of cognitive flexibility -- in 28 healthy individuals during functional neuroimaging. We construct individualized anatomical networks from diffusion spectrum imaging data, allowing us to reconstruct the white matter networks for each subject. We apply a novel approach from \textit{graph signal processing} to distill functional brain signals into \textit{aligned} and \textit{liberal} components relative to underlying anatomy. We show that relatively alignment of brain signals with underlying anatomical network organization supports faster cognitive switching. Moreover, we show that the signals most aligned with the network from moment to moment are also the most flexible. This indicates that anatomical networks provide organization to the flexible brain and facilitate greater mental agility and establishes a potential neural biomarker that integrates brain structure and function.
	
\newpage
			
\section*{Abstract}
Cognitive flexibility describes the human ability to switch between modes of mental function to achieve goals. Mental switching is accompanied by transient changes in brain activity, which must occur atop an anatomical architecture that bridges disparate cortical and subcortical regions by underlying white matter tracts. However, an integrated perspective regarding how white matter networks might constrain brain dynamics during cognitive processes requiring flexibility has remained elusive. To address this challenge, we applied emerging tools from graph signal processing to decompose BOLD signals based on diffusion imaging tractography in 28 individuals performing a perceptual task that probed cognitive flexibility. We found that the \emph{alignment} between functional signals and the architecture of the underlying white matter network was associated with greater cognitive flexibility across subjects. Signals with behaviorally-relevant alignment were concentrated in the basal ganglia and anterior cingulate cortex, consistent with cortico-striatal mechanisms of cognitive flexibility. Importantly, these findings are not accessible to unimodal analyses of functional or anatomical neuroimaging alone. Instead, by taking a generalizable and concise reduction of multimodal neuroimaging data, we uncover an integrated structure-function driver of human behavior.
~\\
~\\

\newpage
\section*{Introduction}
Cognitive flexibility is involved in virtually every complex behavior from mental arithmetic to processing visual stimuli. For example, when navigating complex environments, humans can flexibly switch between two foci of attention or between two processing modalities, in order to effectively respond to sensory inputs. While a hallmark of human cognition, flexible switching is also associated with a measurable cost: moving from one task to another induces a natural extension in the time it takes a person to respond \cite{rogers1995costs}. In patients with neurological syndromes, this cost is even greater, to the point where it can hamper a patient's ability to engage in the basic activities of daily living \cite{szczepanski2014insights}, impacting long-term cognitive outcomes \cite{clark2012specific}. In healthy individuals, cognitive flexibility varies considerably, and individual differences in this trait contribute to mental facets ranging from the development of reasoning ability \cite{richland2013early} to quality of life into late age \cite{davis2010independent}. Clarifying the nature of cognitive flexibility in the human brain is critical to understand the human mind.

The physiological origins of cognitive flexibility are thought to lie in corticobasal ganglia-thalamo-cortical loops \cite{gunaydin2016cortico}: regions of the fronto-parietal and cingulo-opercular systems are activated by cognitive switching tasks \cite{casey2004early, cole2013multi,heyder2004cortico,luk2012cognitive}. In switching paradigms, the anterior cingulate is thought to contribute negative feedback detection following switches, whereas the lateral prefrontal cortex maintains rules and inhibits incorrect responses. Both of these regions anatomically connect to subcortical regions, which are postulated to mediate processes that both suppress prepotent motor responses and transition between behavioral outputs to meet task goals \cite{hikosaka2010switching}. Interactions between cortical systems and motor outputs are thought to be anatomically mediated by subcortical circuits \cite{casey2004early,hosoda2012neural,leunissen2014subcortical,yehene2008basal,heyder2004cortico}. Yet, understanding exactly how this circuit supports task switching has remained challenging, particularly because it requires a conceptual integration of regional activity, inter-regional anatomical connectivity, and observable measures of behavior. While regional activity and behavioral markers of cognitive flexibility are relatively straightforward to estimate, it is less straightforward to integrate these features with the white matter structure (the \emph{connectome} \cite{sporns2005}) that guides the propagation of functional signals \cite{alstott2009, Hermundstad2013,Honey2007}. 

As a result, these two research enterprises -- cognitive neuroscience and connectomics -- have largely developed in parallel without significant cross-talk. Ideally, frameworks that include a concise correspondence between brain network structure, function, and cognitive measures have the potential to produce more comprehensive understanding of human cognition \cite{medaglia2015cognitive,sporns2014challenges}. Conceptually, underlying white matter network organization in the brain physically mediates communication among brain regions. However, analytic frameworks that explicitly use white matter structure to constrain cognitively relevant functional signals are lacking. Such approaches may allow investigators to adjudicate the relative contributions of well-described \emph{systems} in the brain \cite{power2011functional,cole2013multi} to specific cognitive variables by integrating neurophysiological dynamics and anatomy.

\begin{figure}[h!]
	\centerline{\includegraphics[width=3.5in]{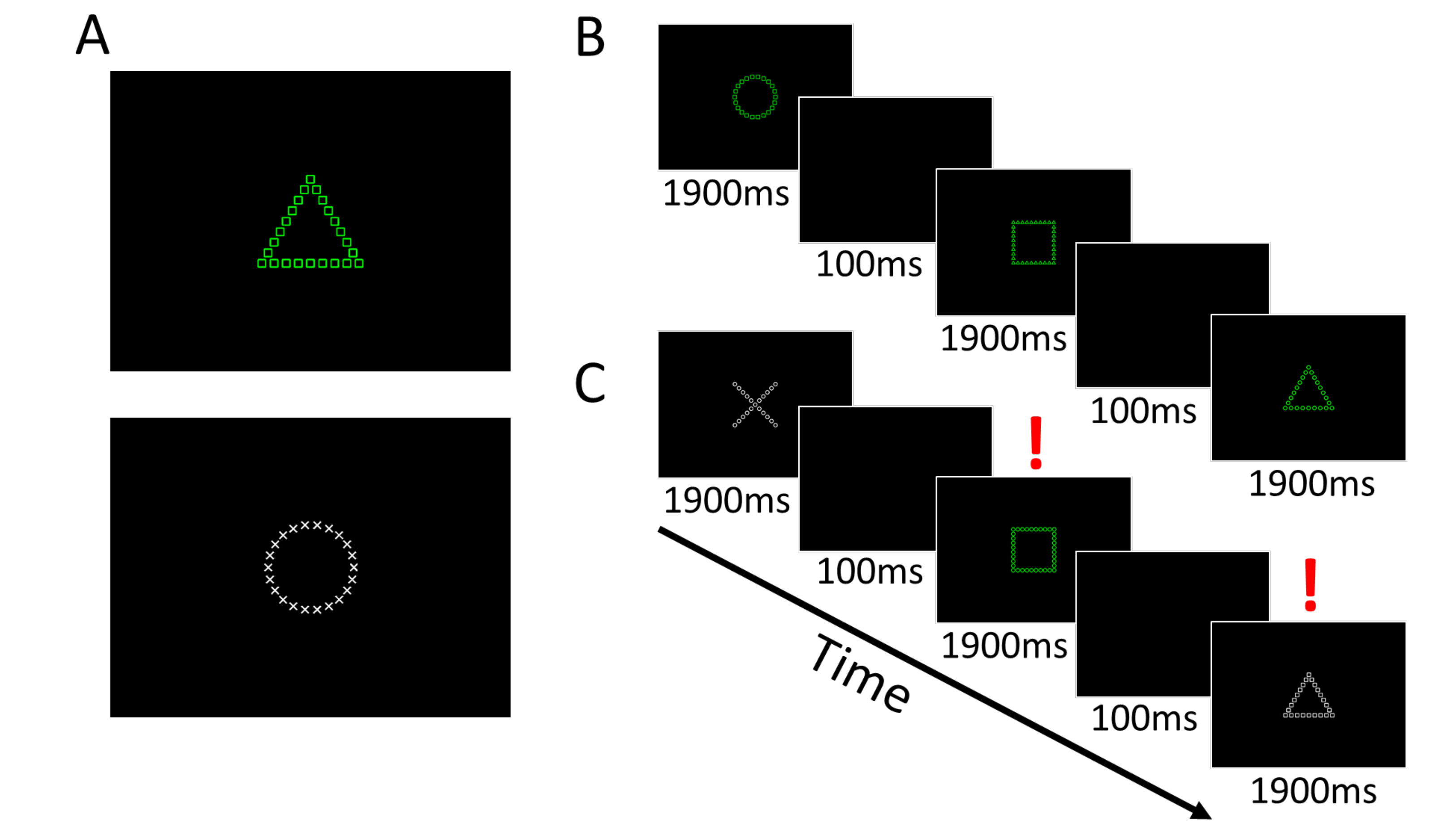}}
	\caption{\textbf{Cognitive task requiring perceptual switching.} \emph{(A)} Example stimuli based on Navon local-global features. Subjects were trained to respond to the larger (or ``global'') shape if the stimulus was green and to the smaller (or ``local") shapes if it was white. \emph{(B)} An example of the non-switching condition for responses. Subjects viewed a sequence of images and were instructed to respond as quickly and accurately as possible. \emph{(C)} An example of the switching condition between stimuli requiring global and local responses. Here, trials with a red exclamation point are switches from the previous stimulus.} \label{fig:task}
\end{figure}

\begin{figure}[h!]
	\centerline{\includegraphics[width=6in]{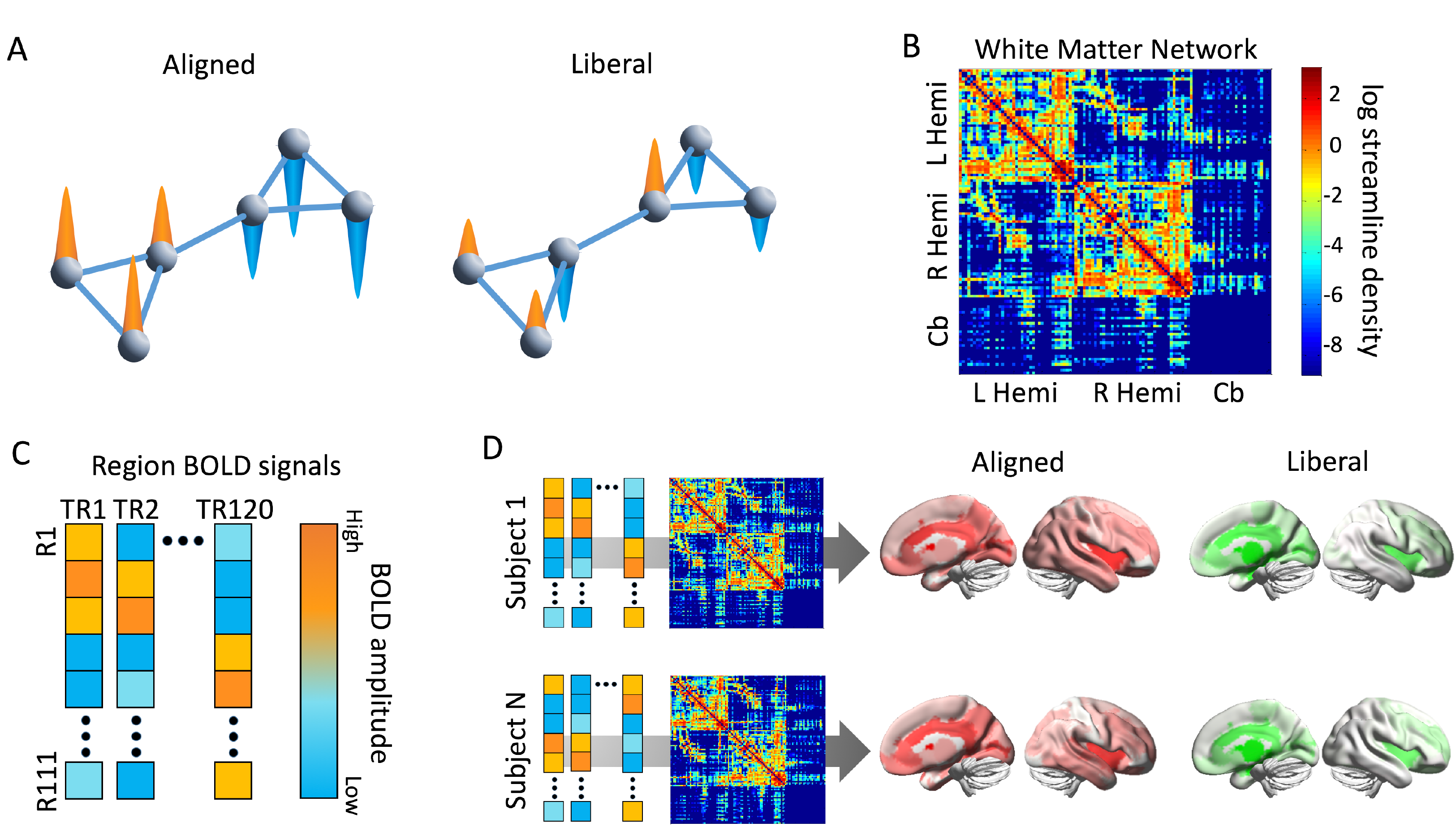}}
	\caption{\textbf{Multimodal approach to the study of cognitive switching using emerging graph signal processing tools.}  \emph{(A)} A notion of signal independence on a schematic modular network. \emph{Left:} An aligned signal on top of a given graph is one in which the magnitude of functional signals, represented by the directionality of the colored cones, corresponds tightly to that expected by the network's organization. In this toy example, one cluster of nodes contains similar positive signals, and the other cluster contains similar negative signals. \emph{Right:} A liberal signal on top of a given graph is one in which signals diverge significantly from the underlying network. \emph{(B)} We construct a white matter graph (adjacency matrix) from 111 anatomically-defined regions where connections are the streamline density between region pairs. \emph{(C)} From BOLD fMRI data acquired during the performance of the Navon task, we extract regional mean time series which we treat as graph signals. \emph{(D)} For each subject, we decompose the graph signals into aligned and liberal components using the underlying eigenspectrum of the white matter graph. We spatially map aligned and liberal signals, and we determine their relationship to switch costs estimated from behavioral performance on the task. Cb = cerebellum. See Methods for details.} \label{methods}
\end{figure}

To address this challenge, we aim to identify the multimodal integration of network anatomy and functional signals that supports cognitive switching. We introduce a novel approach that allows us to examine the distinct contributions of functional signals in the context of anatomically linked regions in human brain networks. In a cohort of 28 healthy adult human subjects, we collected high-resolution diffusion spectrum imaging (DSI) data as well as BOLD fMRI data acquired during the performance of a novel cognitive switching paradigm built on a set of shapes that could be perceived as composed of different features at the local \emph{versus} global scales \cite{navon1977forest} (see Fig.~\ref{fig:task}). From the DSI data, we constructed anatomical brain networks in which 111 cortical, subcortical, and cerebellar regions \cite{cammoun2012,diedrichsen2009probabilistic} were linked pairwise by the density of streamlines reconstructed by state-of-the-art tractography algorithms. Next, we used the eigenspectrum of these anatomical networks to measure the relative separation of framewise regional BOLD signals from the underlying white matter (see Fig.~\ref{methods} and Methods). Specifically, each regional signal was decomposed into a portion that aligned tightly with the anatomical network (\emph{``aligned''}) and a portion that did not align tightly with the network (\emph{``liberal''}). Here, alignment and liberality refer to signal deviations from the underlying anatomical network. We examine the observed BOLD activity for each measurement in time to identify where and to what extent BOLD signals across the brain are organized by white matter networks. Conceptually, this technique allows us to identify to what degree individual BOLD signals deviate weakly \emph{versus} strongly from underlying white matter anatomy.

We anticipate that functional alignment with anatomical networks is an individually variable feature that facilitates cognitive flexibility. We hypothesize that moment-to-moment alignment in human brain networks facilitates switching performance, indicating interindividual variability in the degree of organization of information processing by anatomical topology. This allows us to investigate a theoretical dynamic property supporting cognitive flexibility and provides a mapping to neuroanatomical theories of cognitive switching. In the current study, our task involves \emph{proactive} switching following detection of a contextual cue \cite{hikosaka2010switching}. Whereas prior literature has focused on region-specific mechanisms associated with this process, the current approach allows us to examine the role of \emph{local} neural processes across the brain's \emph{distributed} anatomical network. We postulate that functional alignment serves as a mechanism for cognitive flexibility across the brain's complex anatomical white matter network. In particular, well-aligned signals may represent efficiently organized signals with markedly less interference, whereas liberal signals do not leverage the inherent organization of underlying white matter networks. This approach to multimodal neuroimaging could distinguish whether structure and function operate synergistically or in opposition to promote cognitive flexibility. Accordingly, we examine whether alignment with anatomically mediated expectations in subcortical regions, cortical systems, or both constitutes the basis for effective cognitive switching. This allows us to determine (i) whether alignment forms a basis for cognitive flexibility and (ii) whether regions such as those in the subcortical system that serve an \emph{anatomically} central network role also serve a greater or less \emph{functionally} central role in this mental skill.

\newpage

\section*{Functional alignment in liberal subcortical systems is associated with faster cognitive switching}

\begin{figure}[h!]
	\centerline{\includegraphics[width=3.5in]{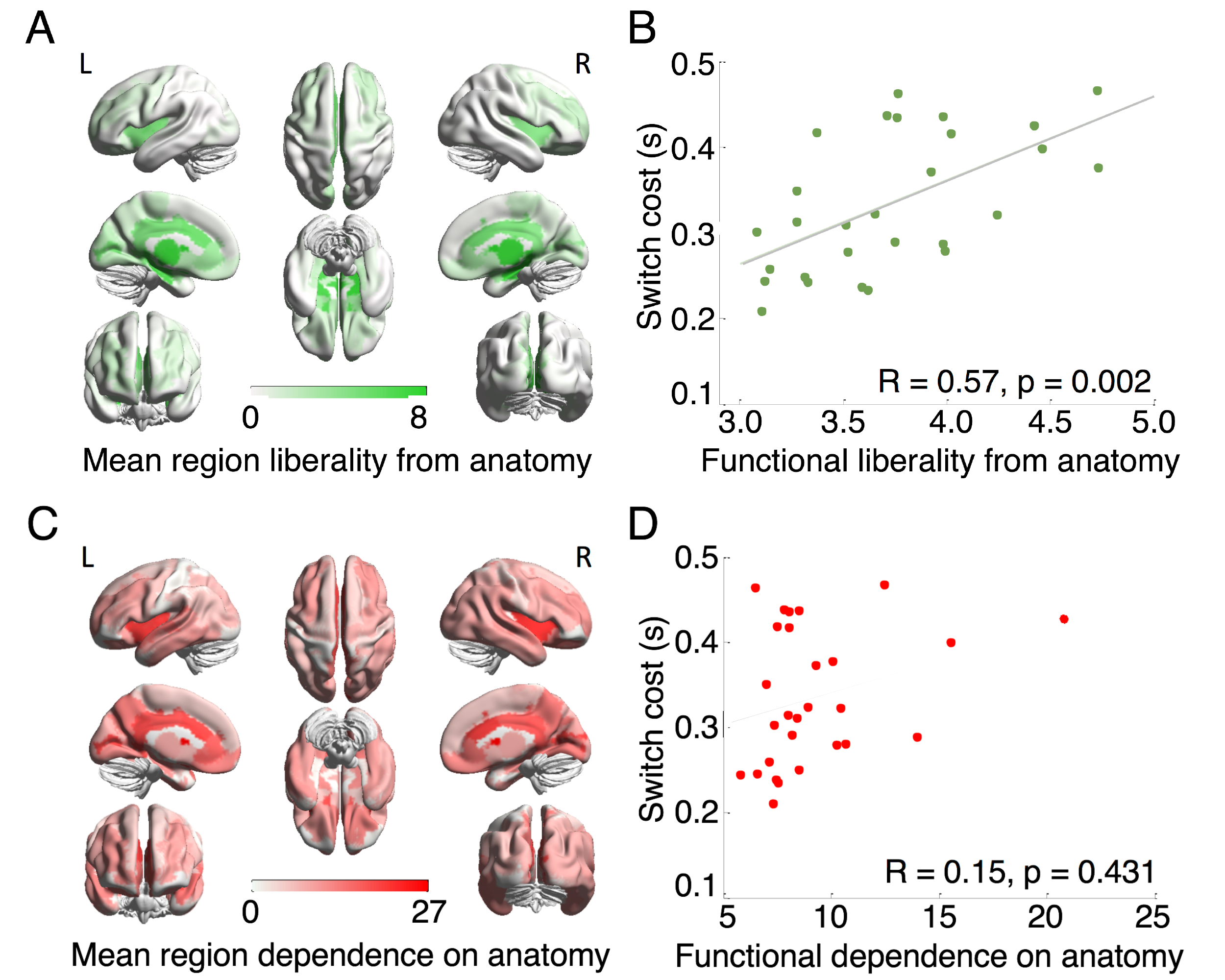}}
	\caption{\textbf{Lower independence is associated with lower switch costs.}  \emph{(A)} Liberal signals were concentrated especially in subcortical regions and cingulate cortices. \emph{(B)} Reduced liberality (increased alignment) is associated with reduced switch costs across subjects. \emph{(C)} Aligned signals were concentrated especially in subcortical, default mode, fronto-parietal, and cingulo-opercular systems. \emph{(D)} Variability in aligned signals was not significantly associated with switch costs across subjects. L = left hemisphere, R = right hemisphere.} \label{independenceshift}
\end{figure}

\begin{figure}
	\centerline{\includegraphics[width=3.5in]{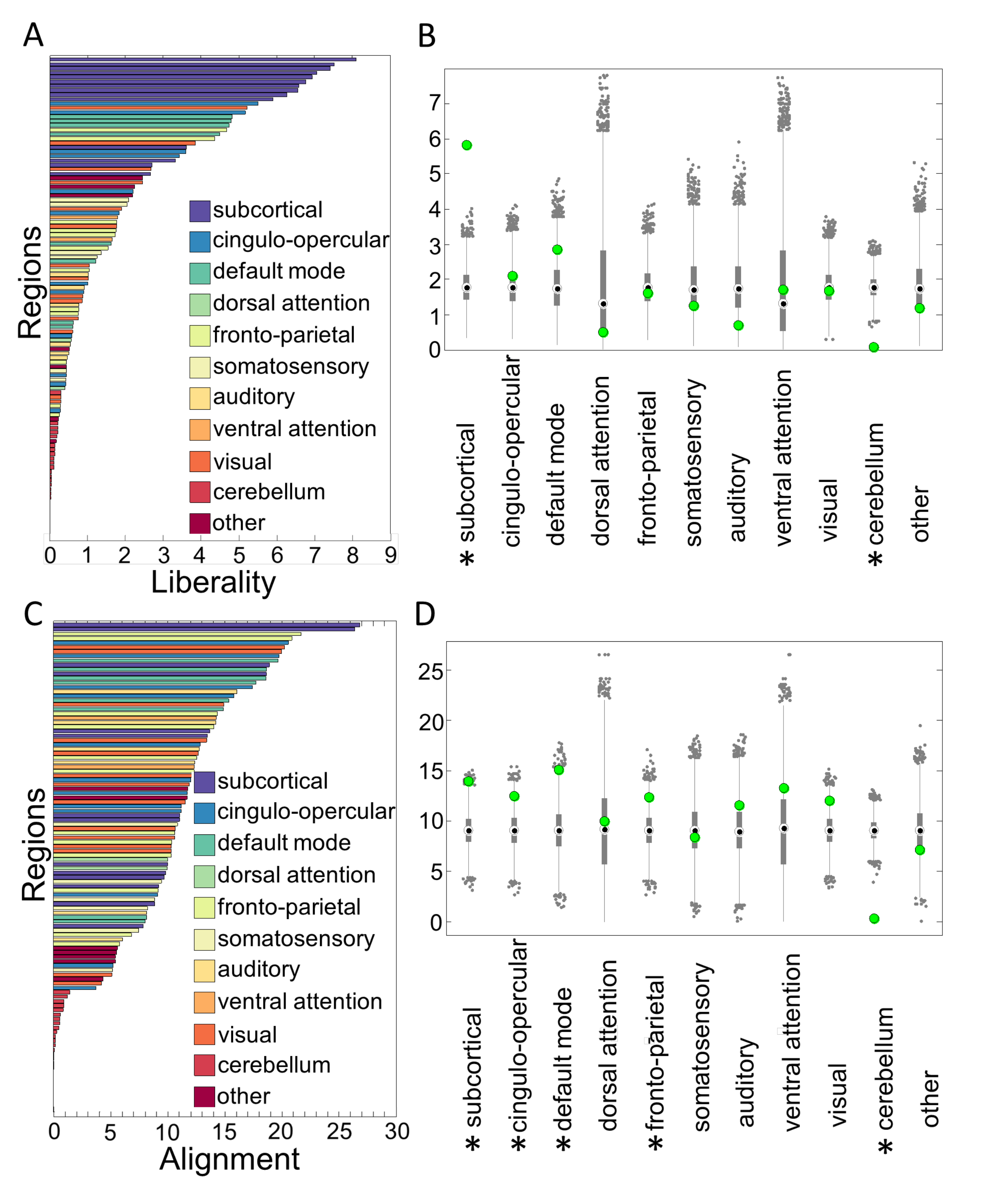}}
	\caption{\textbf{Non-parametric permutation test for signal concentration within cognitive systems.} \emph{(A)} Liberal signal concentrations sorted from highest (\emph{top}) to lowest concentration across all regions in all subjects. \emph{(B)} Liberal signals were most concentrated in subcortical regions. \emph{(C)} Aligned signal concentrations sorted from highest (\emph{top}) to lowest concentration across all regions in all subjects. \emph{(D)} Aligned signals were most concentrated in fronto-parietal, cingulo-opercular, default mode, and subcortical systems.} \label{systemconcentration}
\end{figure}

Aligned signals were concentrated within default mode, fronto-parietal, cingulo-opercular, and subcortical systems across subjects, whereas the liberal signals were concentrated largely in the subcortical system (Fig.~\ref{systemconcentration}). The significance of these concentrations was confirmed statistically using a non-parametric permutation test ($\alpha=0.05$) in which we shuffled the values of alignment (or liberality) uniformly at random across brain regions before computing the mean alignment (or liberality) value within each system \cite{Gu2015}. Then, we calculated the correlation between aligned and liberal BOLD signals across the brain and cognitive switch costs (response times during trials with a color-cued switch \emph{versus} non-switching trials).  We observed that variability in \emph{aligned} signals was not associated with switch costs ($R = 0.15, p = 0.43$, accounting for 2\% of the variance), while variability in \emph{liberal} signals accounted for 32\% of variance (see Fig.~\ref{independenceshift}). Among the liberal signals, \emph{lower} values of liberality (that is, relative alignment) were also associated with lower switch costs both during fixation ($R = 0.62, p = 0.0006$) and during nonswitching ($R = 0.71, p = 0.0001$) perceptual blocks.

It is interesting to note that signals extracted from subcortical areas can be decomposed into both highly anatomically-aligned portions and highly liberal portions (see Methods). The values of both alignment and liberality were significantly greater than expected in subcortical structures. Yet, other systems show a disposition toward high alignment (fronto-parietal, cingulo-opercular, and default mode systems), or no disposition toward alignment or liberality (visual, auditory, and dorsal \& ventral attention). We quantified the relationship between aligned and liberal signals by calculating the Pearson correlation coefficient between alignment and liberality values over all brain regions. We observed that the two types of signals were significantly but not perfectly correlated across the brain: mean over subjects and tasks $R = 0.69, p \ll 0.001$ (see also Supplement). This analysis highlights an interesting measure of multimodal complexity that differs across brain areas: subcortical regions can be thought of as displaying high multimodal complexity because they contain both highly anatomically-aligned signals and highly liberal signals.

\section*{Anatomy and function together define a neural trait underlying cognitive flexibility}

Finally, we asked whether the observed regional variation in aligned and liberal signals could be estimated reliably across subjects and tasks, supporting its utility as a marker for behavior. To address this question, we compared the regional patterns of alignment between subjects and between task conditions (fixation, non-switching, and switching). We observed that the regional pattern of signal alignment was consistent across the three task conditions for a given subject (intraclass correlation: $R = 0.99, p \ll 0.001$), as was the regional pattern of liberal signals (intraclass correlation: $R = 0.99, p \ll 0.001$; see also Supplement). These results indicate that regional patterns of signal alignment and liberality are highly stable phenotypes of a subject across the three task conditions.
 
Taken together, our results demonstrate that individuals whose most liberal functional signals were more aligned with white matter architecture could switch perceptual focus faster. In other words, relative alignment with anatomy was associated with greater cognitive flexibility, a finding that highlights the importance of simultaneously considering both functional and anatomical neuroimaging in the study of higher order cognitive processes. These findings complement prior studies of executive function that have focused on node-level, edge-level, and module-level features of brain networks \cite{braun2015dynamic,leunissen2014subcortical} (see also Supplement). Here, we consider brain function as a series of time-evolving states \cite{Gu2015,mayhew2016global} that are organized in relation to the underlying pattern of white matter tracts. The state-based focus of our approach also offers insights into the differential extent to which specific cognitive systems deviate from tract anatomy, underscoring anatomical contributions to the organization of brain dynamics across subjects.

Our results contextualize previous models of cortico-striatal cognitive switching mechanisms \cite{hikosaka2010switching,sekutowicz2016striatal} within a connectomic perspective. As a complement to prior findings implicating individual prefrontal and striatal systems in cognitive switching, our results highlight the importance of anatomical network organization, and the central role of subcortical functional dynamics atop that structure. Specifically, we found that modest functional alignment of the most liberal signals in the subcortical and cingulate cortices was associated with faster cognitive switching. This observation is particularly interesting in the context of prior work showing that subcortical and anterior cingulate regions manage multiple inputs and outputs among cortical systems during task transitions \cite{liston2006anterior,heyder2004cortico}, potentially requiring more diverse signal organization relative to anatomical networks. Interestingly, we observe that the levels of alignment and liberality were consistent within subjects across all three conditions of the cognitive task -- fixation, non-switching, and switching blocks -- suggesting that it may be a trait rather than a state of an individual. While prior studies have similarly identified consistent network architecture across fixation and task \cite{cole2014intrinsic}, these similarities have not previously been identified in a model-based integration of functional dynamics and white matter anatomy. The observed inter-task consistency of functional signal alignment and liberality was complemented by broad inter-subject variability. We speculate that this variability represents differences in the amount of energy utilized to enact perceptual switches \cite{watanabe2014energy}, with greater independence requiring greater energy. Thus, future studies could examine the potentially subject-specific role of cognitive control systems in regulating energy-efficient state transitions in local-global processing.

With respect to recent dynamic network analyses of executive function, this analysis contributes a crucial anatomically-grounded perspective. The current approach represents a framework in which to understand the dual features of anatomical organization and functional processes supporting cognitive flexibility in the human connectome. Here, high functional dependence in fronto-parietal, cingulo-opercular, default mode, and subcortical systems is not associated with intersubject switching variability. Critically, our results indicate that regions that participate in highly flexible systems \cite{mattar2016flexible} in \emph{temporal network analysis} demonstrate high dependence on underlying anatomical networks across frames of BOLD data during fixation, low cognitive control conditions, and task conditions (see Supplement for further analysis and discussion). Previous studies identify dynamic network roles for fronto-parietal and cingulo-opercular regions in cognitive switching, and our results indicate that moment to moment signal configurations in highly flexible systems are strongly organized by structure across time (see Supplement for addition analysis and discussion). In the context of this highly organized cortical activity, the current results suggest that subcortical systems contain highly liberal signals. The extent to which subcortical systems exhibit relative alignment may form a flexible integrative role across the many computations supported by cortical systems. The relationship between anatomically-bound momentary signal organization and functional reconfigurations in temporal networks may more generally provide a fruitful area for future research.

Notably, our results do not explain the potential cognitive role of highly aligned signals. It is possible that the role of these signals may be better explained in the context of other cognitive control processes \cite{Miyake2000}. We speculate that anatomically aligned signals in fronto-parietal, cingulo-opercular, default mode, and subcortical systems organize the dynamic signals contributing to cortical mechanisms of cognitive control, attention, and resting and preparedness processes, respectively. It would be interesting to test whether highly aligned signals in association cortices and subcortical structures are associated with domain-general performance variability across modalities \cite{fedorenko2014role}. In cognitive switching specifically, the extent to which signal liberality relates to performance on tasks involving other sensory modalities, transitioning between internally and externally focused attention, and divergent thinking remains to be established.

In conclusion, our study supports the utility of network science in clarifying mechanisms of executive function specifically and cognition more generally \cite{medaglia2015cognitive, sporns2014challenges}. Recent literature firmly establishes that white matter organization is a critical, but incomplete determinant of functional signals in brain networks \cite{Hermundstad2013,hermundstad2014structurally}. Conceptually, the current approach acknowledges that without structure, functional signals lack a mediating organization. By examining functional signal \emph{alignment} within underlying white matter networks, we identify an important definition of dynamic contributions to cognitive switching that powerfully discriminates between the contributions of subcortical and other systems in the brain. Similar applications to other large multimodal neuroimaging datasets could contribute to biomarker analyses in psychiatric disease and neurological disorders, many of which are associated with deficits in executive function \cite{belleville2008task,kehagia2010neuropsychological,kinnunen2010white}.

\newpage

\section*{Methods}
\subsection*{Subjects}
 A total of 30 subjects were recruited. All subjects were screened for prior history of psychiatric or neurological illness. One subject was excluded due to near-chance performance on the task (accuracy = 52\%). One additional subject was excluded due to technical problems on the day of scanning. The final sample included 28 individuals (mean age = 25.6, St.D. = 3.5, 70\% caucasian, 13 females). All subjects volunteered with informed consent in writing in accordance with the Institutional Review Board/Human Subjects Committee, University of Pennsylvania.
 
 \subsection*{Behavioral task}
All participants completed a local-global perception task based on classical Navon figures \cite{navon1977forest}. Local-global stimuli were comprised of four shapes -- a circle, X, triangle, or square -- that were used to build the global and local aspects of the stimuli. On all trials, the local feature did not match the global feature, ensuring that subjects could not use information about one scale to infer information about another. Stimuli were presented on a black background in a block design with three block types (See Fig.~2). In the first block type, subjects viewed white local-global stimuli. In the second block type, subjects viewed green local-global stimuli. In the third block type, stimuli switched between white and green across trials uniformly at random with the constraint that 70\% of trials included a switch in each block. In all blocks, subjects were instructed to report only the local features of the stimuli if the stimulus was white and to report only the global feature of the stimuli if the stimulus was green. Blocks were administered in a random order. Subjects responded using their right hand with a four-button box. All subjects were trained on the task outside the scanner until proficient at reporting responses using a fixed mapping between the shape and button presses (i.e., index finger = ``circle'', middle finger = ``X'', ring finger = ``triangle", pinky finger = ``square''). In the scanner, blocks were administered with 20 trials apiece separated by 20 s fixation periods with a white crosshair at the center of the screen. Each trial was presented for a fixed duration of 1900 ms separated by an interstimulus interval of 100 ms during which a black screen was presented.

\subsection*{Diffusion spectrum imaging acquisition and processing}~
Diffusion spectrum images (DSI) were acquired on a Siemens 3.0T Tim Trio for all subjects along with a T1-weighted anatomical scan at each scanning session. We followed a parallel strategy for data acquisition and construction of streamline adjacency matrices as in previous work \cite{Gu2015,betzel2016optimally}. DSI scans sampled 257 directions using a Q5 half-shell acquisition scheme with a maximum $b$-value of 5,000 and an isotropic voxel size of 2.4 mm. We utilized an axial acquisition with the following parameters: repetition time (TR) = 5 s, echo time (TE)= 138 ms, 52 slices, field of view (FoV) (231, 231, 125 mm). We acquired a three-dimensional SPGR T1 volume (TE = minimal full; flip angle = 15 degrees; FOV = 24 cm) for anatomical reconstruction. All subjects volunteered with informed consent in writing in accordance with the Institutional Review Board/Human Subjects Committee, University of Pennsylvania.

DSI data were reconstructed in DSI Studio (www.dsi-studio.labsolver.org) using $q$-space diffeomorphic reconstruction (QSDR)\cite{yeh2011estimation}. QSDR first reconstructs diffusion-weighted images in native space and computes the quantitative anisotropy (QA) in each voxel. These QA values are used to warp the brain to a template QA volume in Montreal Neurological Institute (MNI) space using the statistical parametric mapping (SPM) nonlinear registration algorithm. Once in MNI space, spin density functions were again reconstructed with a mean diffusion distance of 1.25 mm using three fiber orientations per voxel. Fiber tracking was performed in DSI studio with an angular cutoff of 35$^\circ$, step size of 1.0 mm, minimum length of 10 mm, spin density function smoothing of 0.0, maximum length of 400 mm and a QA threshold determined by DWI signal in the colony-stimulating factor. Deterministic fiber tracking using a modified FACT algorithm was performed until 1,000,000 streamlines were reconstructed for each individual.

Anatomical scans were segmented using FreeSurfer\cite{fischl2012freesurfer} and parcellated using the connectome mapping toolkit \cite{cammoun2012mapping}. A parcellation scheme including $n=129$ regions was registered to the B0 volume from each subject's DSI data. The B0 to MNI voxel mapping produced via QSDR was used to map region labels from native space to MNI coordinates. To extend region labels through the grey-white matter interface, the atlas was dilated by 4 mm \cite{cieslak2014local}. Dilation was accomplished by filling non-labelled voxels with the statistical mode of their neighbors' labels. In the event of a tie, one of the modes was arbitrarily selected. Each streamline was labeled according to its terminal region pair.

Finally, we included a cerebellar parcellation \cite{diedrichsen2009probabilistic}. We used FSL to nonlinearly register the individual's T1 to MNI space. Then, we used the inverse warp parameters to warp the cerebellum atlas to the individual T1. We registered the subject's DSI image to the T1. We used the inverse parameters from this registration to map the individualized cerebellar parcels into the subject's DSI space. Finally, we merged the cerebellar label image with the dilated cortical and subcortical parcellation image.

From these data and parcellation, we constructed an anatomical connectivity matrix, $\mathbf{A}$ whose element $A_{ij}$  represented the number of streamlines connecting different regions \cite{Hermundstad2013}, divided by the sum of volumes for regions $i$ and $j$ \cite{hagmann2008mapping}. Prior to data analysis, all cerebellum-to-cerebellum edges were removed from each individual's matrix because cerebellar lobules are demonstrably not anatomically connected directly to one another \cite{voogd1998anatomy}.

\subsection*{Functional imaging acquisition and processing}~
fMRI images were acquired during the same scanning session as the DSI data on a 3.0T Siemens Tim Trio whole-body scanner with a whole-head elliptical coil by means of a single-shot gradient-echo T2* (TR = 1500 ms; TE = 30 ms; flip angle = 60$^{\circ}$; FOV = 19.2 cm, resolution 3mm x 3mm x 3mm). Preprocessing was performed using FEAT v. 6.0 (fMRI Expert Analysis Tool) a component of the FSL software package \cite{jenkinson2012fsl}. To prepare the functional images for analyses, we completed the following steps: skull-stripping with BET to remove non-brain material, motion correction with MCFLIRT (FMRIB's Linear Image Registration Tool; \cite{jenkinson2012fsl}), slice timing correction (interleaved), spatial smoothing with a 6-mm 3D Gaussian kernel, and high pass temporal filtering to reduce low frequency artifacts. We also performed EPI unwarping with fieldmaps in order to improve subject registration to standard space. Native image transformation to a standard template was completed using FSL's affine registration tool, FLIRT \cite{jenkinson2012fsl}. Subject-specific functional images were co-registered to their corresponding high-resolution anatomical images via a Boundary Based Registration technique (BBR \cite{greve2009accurate}) and were then registered to the standard MNI-152 structural template via a 12-parameter linear transformation. Finally, each participant's individual anatomical image was segmented into grey matter, white matter, and CSF using the binary segmentation function of FAST v. 4.0 (FMRIB's Automated Segmentation Tool \cite{zhang2001segmentation}). The white matter and CSF masks for each participant were then transformed to native functional space and the average timeseries were extracted. Images were spatially smoothed using a kernel with a full-width at half-maximum of 6 mm. These values were used as confound regressors on our time series along with 18 translation and rotation parameters as estimated by MCFLIRT \cite{Jenkinson2002}.

\subsection*{Functional decomposition into anatomical networks}
We analyze the signal defined on a connected, weighted, and symmetric graph, $\ccalG=(\ccalV, \bbA)$, where $\ccalV=\{1,\ldots,n\}$ is a set of $n$ vertices or nodes representing individual brain regions and $\bbA \in \reals^{n \times n}$ is defined as above. Because the network $\bbA$ is symmetric, it has a complete set of orthonormal eigenvectors associated with it \cite{Chung97, sandryhaila2013discrete}. For this reason, it has an eigenvector decomposition, $\bbA =  \bbV \bbLambda \bbV^T$, in which $\bbLambda$ is the set of eigenvalues, ordered so that $\lam_0\leq\lam_1\leq\ldots\leq\lam_{n-1}$, and $\bbV = \{\bbv_k\}_{k = 0}^{n-1}$ is the set of associated eigenvectors. Following \cite{sandryhaila2013discrete, shuman2013}, we use the eigenvector matrix to define the Graph Fourier Transform (GFT) of the graph signal $\bbx \in \reals^n$, defined as
\vspace{-0.5mm}
\begin{align}\label{eqn_GFT}
	\tbx = \bbV^T \bbx .
\end{align}
\vspace{-1mm}

\noindent Given $\tbx=[\tdx_0,\ldots,\tdx_{n-1}]^T$, we can express our original signal as $\bbx = \sum_{k=0}^{n-1} \tdx_k \bbv_k$, a sum of the eigenvector components $\bbv_k$. The contribution of $\bbv_k$ to the signal $\bbx$ is the GFT component $\tdx_k$. Note that the smoothness of $\bbv_k$ on the network can be evaluated in the quadratic form $\bbv_k^T \bbA \bbv_k = \sum_{i, j \in V} A_{ij} v_k(i) v_k(j)$ and that $\bbv_k^T \bbA \bbv_k = \lambda_k$ is given by the eigenvector decomposition. The quantity $\sum_{i, j \in V} A_{ij} v_k(i) v_k(j)$ will be negative when the signal is varied (highly connected regions possess signals of different signs), and positive when the signal is smooth (highly connected regions possess signals of same signs); for these reasons, this quantity can be thought of as a measure of smoothness (\textbf{alignment}). Consequently, these GFT coefficients $\tdx_k$ for small values of $k$ indicate how much variables that are highly misaligned (\textbf{liberality}) with structure contribute to the observed brain signal $\bbx$. GFT coefficients $\tdx_k$ for large values of $k$ describe how much signals that are aligned with the anatomical network contribute to the observed brain signal $\bbx$. The inverse (i)GFT of $\tbx$ with respect to $\bbA$ is defined as 
\vspace{-0.5mm}
\begin{align}\label{eqn_iGFT}
	\bbx = \bbV \tbx.
\end{align}
\vspace{-1mm}

Given a graph signal $\bbx$ with GFT $\tbx$, we can isolate the liberal components corresponding to the lowest $K_\L$ eigenvectors by applying a graph filter $\bbH_\L$ that only keeps components with $k < K_\L$ and sets other components to $0$. The signal $\bbx_\L$ then contains the ``liberal'' components of $\bbx$ (those with a \emph{low} alignment with network structure). Apart from the graph low-pass filter $\bbH_\L$, we also consider a middle graph regime $\bbH_\M$, which keeps only components in the range of $K_\L \leq k < n - K_\A$, and an ``aligned'' graph regime $\bbH_\A$, such that only network-aligned components with $n - K_\A \leq k$ are kept. Therefore, the liberal regime takes the lowest $K_\L$ components, the alignment regime takes the highest $K_\A$ components, and the middle regime captures the middle $n - K_\L - K_\M$ components (here, we use the components with the $10$ lowest alignment values to represent the liberal regime and the components with the $10$ highest alignment values to represent the aligned regime; see Supplement for robustness of results to parameter selection). As such, if we use $\bbx_\M$ and $\bbx_\A$ to respectively denote the signals represented by the middle and highly aligned regimes, we have that the original signal can be written as the sum $\bbx = \bbx_\L + \bbx_\M + \bbx_\A$. This formulation gives a decomposition of the original signals $\bbx$ into liberal, moderately aligned, and highly aligned components that respectively represent signals that have high, medium, and low signal deviation with respect to the anatomical connectivity between brain regions. 

Prior work has consistently demonstrated that the aligned and liberal components aid in better estimation of unknown movie ratings in recommendation systems \cite{ma2015}, better prediction of cancer using gene interaction networks \cite{segarra2015, huang2015diffusion}, and learning in neuroimaging data, where learning-related processes are demonstrably expressed in low and high components in fMRI data, and where the middle component $\bbx_\M$ is demonstrably less reliable and behaviorally uninformative \cite{huang2016graph}. In the supplement, we perform a similar analysis to \cite{huang2016graph} but with the current data to examine the stability of our low and high alignment measurements to parameter selection. The data indicates that the low and high alignment components in the current data are stable. Mathematically, this is expected in general in applications of the current approach because eigenvalues at the extreme low and high end are isolated from the middle values, which leads to robustness in the high and low ranges of the decomposition \cite{spielman2009spectral}.

\begin{figure}
	\centerline{\includegraphics[width=3.5in]{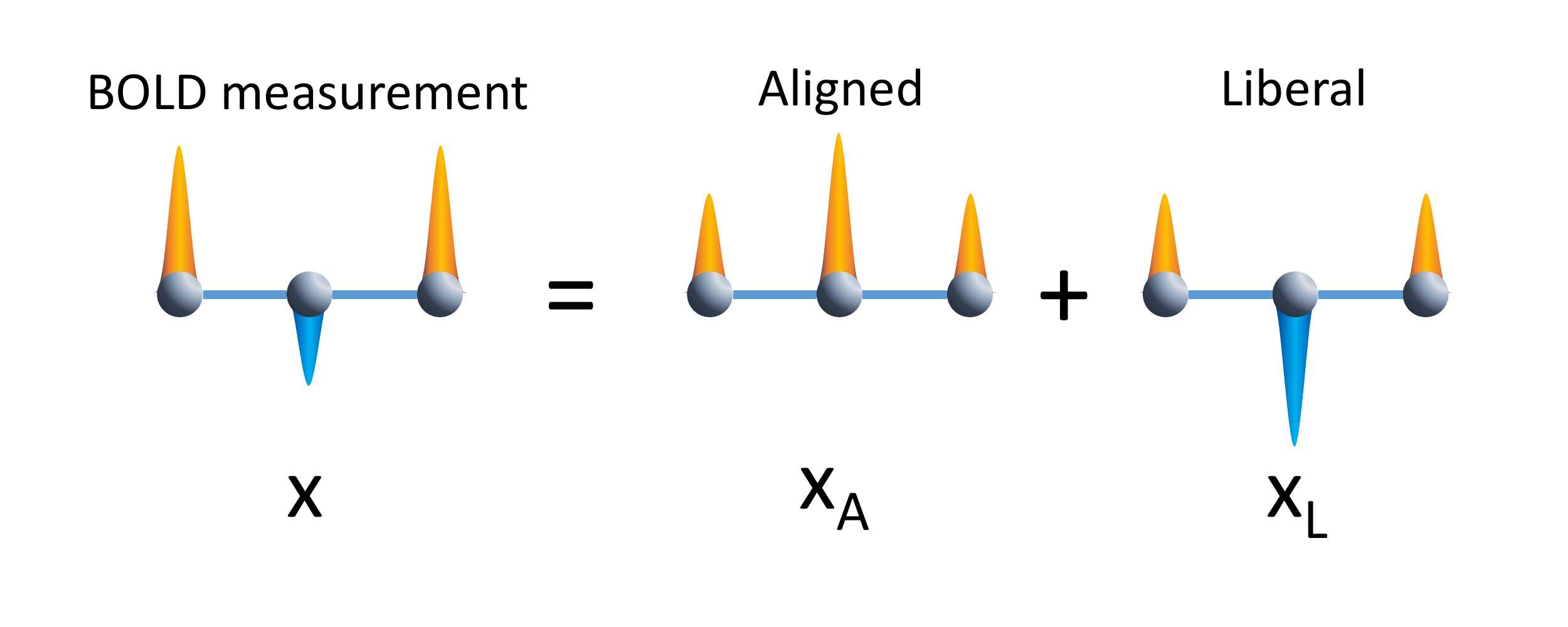}}
	\caption{\textbf{Signal decomposition into anatomy.} BOLD signals are decomposed into aligned and liberal signal components. Left of equation: a schematic BOLD signal on a simple anatomical network. Here, two signals are stronger in the high direction than the low direction. Right of equation: the signals across the network are decomposed into an aligned and liberal component. The original signals can be reconstructed from a basis set including a weighted part of the signal that is aligned with the anatomical network and another part that is liberal with respect to the anatomical network.} \label{fig:GFTintuition}
\end{figure}

We note that this approach allows signals on the anatomical network to contain \textit{both} aligned and liberal components represented in the same region at a single TR. This flexibility occurs because the anatomical network of $n$ nodes has $n^2$ entries (i.e., connection information is encoded in the anatomical adjacency matrix for any node $i$ to any node $j$). Rather than examining a single BOLD signal measurement as $n$ independently observed values, the GFT considers the signal to be a composite of contributions to the signal across subject's anatomical network topology. The decomposition occurs across the entire set of signals (here, the vector of BOLD magnitudes across regions at a single TR), where there are only $n$ entries. The GFT applied here leverages the fact that the $n$ entries in a given vector are not isolated, but are signals on top of the complex anatomical network. In the current approach, instead of focusing on the single BOLD value observed at each region as a discrete entity, the decomposition is sensitive to the observation of pairwise differences among BOLD signals relative to that expected by the anatomical network. Some portion of each given region's BOLD signal is estimated to be liberal with respect to the network, which is represented by $\bbx_\L$, and some portion is estimated to be aligned with the network, which is represented by $\bbx_\A$ (See Fig.~\ref{fig:GFTintuition}). This mathematical separation establishes the notions of alignment and liberality of the BOLD signals in the anatomical network. All individual regions in the brain will have some degree of alignment and some degree of liberality given the complexity of BOLD signal patterns across the network, unless the observed BOLD signals in all regions are perfectly aligned or perfectly misaligned with the subject's anatomical network. This highlights an important strength of the use of the Graph Fourier Transform to examine functional signal liberality in anatomical brain networks: the signal can be understood as a network level composite of aligned and liberal signals, and the extent to which individual regions contribute to these properties can be examined as the variation in the weights of the region's contribution to each of the aligned and liberal components.

\newpage

\subsection*{Relating signals to behavior}
Following the signal decomposition into aligned and liberal signals, we associated the signal concentrations with median switch cost (median response time during switching trials versus no-switching trials) performance for all accurate trials. To do so, we computed a partial Pearson's correlation between the observed signal value for each subject with their median switch cost using the average framewise displacement across BOLD measurements as a second-level control for the influences of motion. Specifically, to examine the relationship between alignment and switch costs across subjects, we computed the partial correlation for the mean of $\bbx_\A$ for each subject with subjects' switch costs, controlling for average framewise displacement. Then, to examine the relationship between liberality and switch costs, across subjects, we computed the partial correlation for the mean of $\bbx_\L$ for each subject with subjects' switch costs, controlling for average framewise displacement. We additionally repeated these analyses including age and sex and found similar slopes of the associations between the liberality values and switch costs (see Supplement). 

\subsection*{System permutation test}~
To examine the spatial significance of system-level concentration of aligned and liberal signals, we performed a non-parametric permutation test for each signal class. Separately for each of $\bbx_\L$ and $\bbx_\A$, we shuffled the observed mean signal concentration values across regions in 10,000 permutations for aligned and liberal signals and computed a null distribution of system mean signal concentrations for each system. Signals were judged to be significantly concentrated in a system if the mean signal concentration in the system was greater or less than 95\% of the null permutations.

\newpage

\section*{Supplementary Information and Analyses}

\subsection*{Behavior}

Across the 28 subjects, individuals were 94\% (St.D. = 1\%) accurate across all trials. The average median response time for accurate non-switching trials was 0.89 s (St.D. = 0.12 s). The average median response time for accurate switching trials across subjects was 1.22 s (St.D. = 0.18 s). The average switch cost across subjects was 0.32 s (St.D. = 0.08 s).

\subsection*{Signal alignment as a trait-level variable in human brain networks}

In the main text, we report intraclass correlations that indicate that alignment and liberality form ``trait'' variables due to the consistency of aligned and liberal signal patterns regardless of the task period during which the signals were sampled. To examine the stability of aligned and liberal signals across all individual conditions, we compute a Pearson's correlation coefficient between the signal concentration vector across regions for each condition pair among aligned and liberal signal vectors for the fixation, no switch, and switching blocks. We observe that the aligned and liberal signals are very highly stable for each pairwise correlation within each signal type, but only moderately correlated across signal types (within aligned: mean $R = 0.99, p \ll 0.001$, within liberal: mean $R = 0.99, p \ll 0.001$, between aligned and liberal: mean $R = 0.69, p \ll 0.001$; See. Fig. 1). These results establishe that during periods of resting fixation, no switching demands, and switching demands, alignment and liberality are highly stable subject-level traits.

\begin{figure} [ht!]
	\centerline{\includegraphics[width=3.5in]{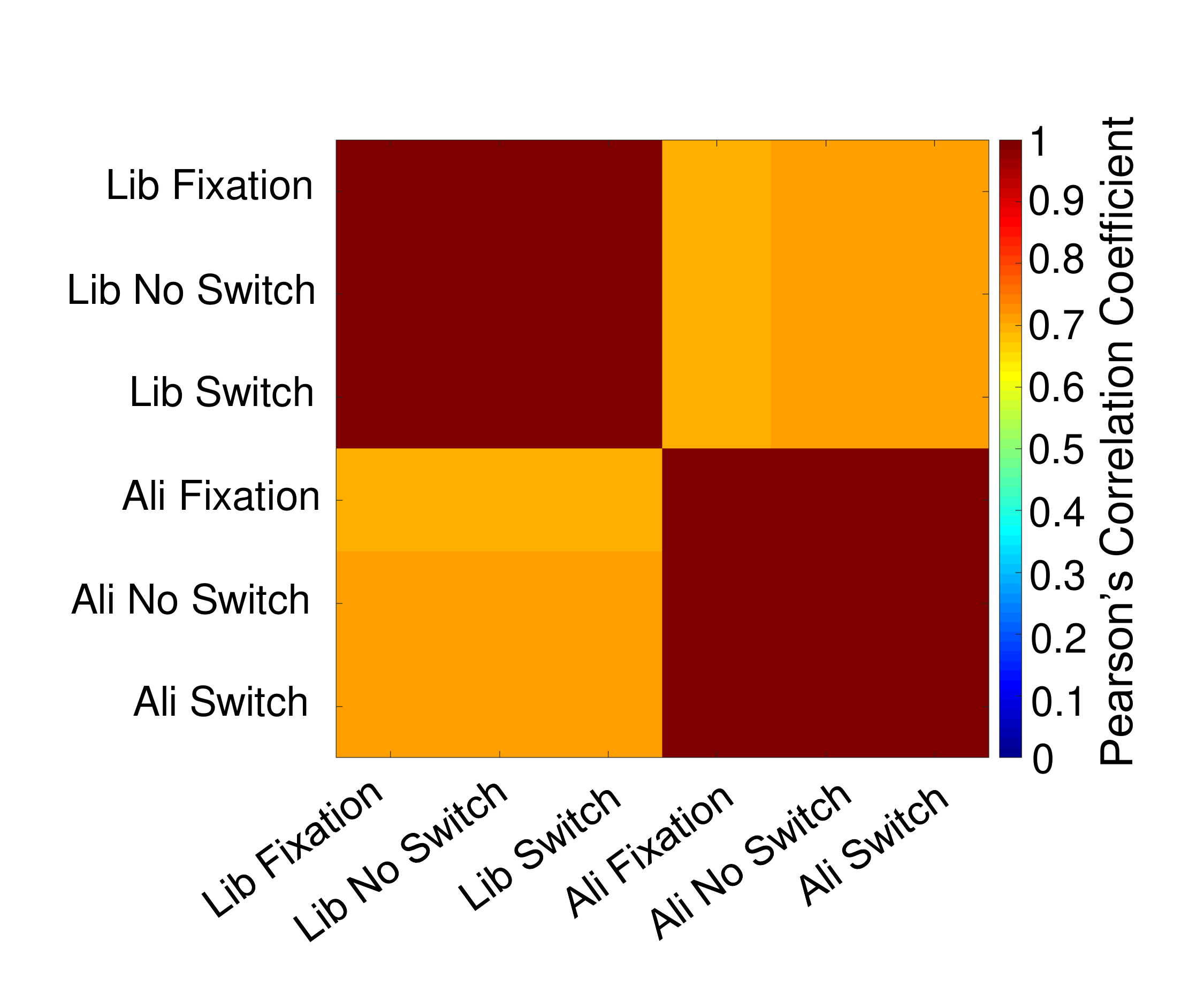}}
	\caption{\textbf{Aligned and liberal signals are consistent traits in human brain networks.} Correlations among conditions decomposing BOLD fMRI signals using the eigenspectrum of individuals' anatomical networks. Aligned and liberal signals are highly stable at the subject level. Ali = Aligned, Lib = Liberal.}\label{task}
\end{figure}

\subsection*{Signal variability in systems}

In the main text, we note that signals that are highly aligned with the underlying anatomical networks were concentrated in fronto-parietal, cingulo-opercular, default mode, and subcortical systems. This is especially interesting given numerous findings identifying the fronto-parietal and cingulo-opercular systems to contain variable signals over time and more flexible changes in network organization during both rest and task performance \cite{mattar2016flexible,braun2015dynamic}. To examine the BOLD signal variability in systems across TRs in the current study, we compute the average standard deviation of BOLD time series values over trials for all regions within each system. Then, we conduct a non-parametric permutation test in which we shuffle the assignment of regions to systems ($n = 10,000$) and compute a null distribution of average time series standard deviations over TRs for each system. We judge a system to be significantly variable or invariant over time if its observed average standard deviation is greater than or less than 97.5\% of null permutations. We find that the fronto-parietal and cingulo-opercular systems are more variable across TRs than the null expectation, whereas the ventral attention, somatosensory, and cerebellar systems are less variable. Considered together with our main results examining aligned and liberal signal concentrations, these results indicate that in cognitive control systems that are functionally dynamic, signal variance is strongly organized by aligning with subject's anatomical networks from TR to TR.

\begin{figure} [ht!]
	\centerline{\includegraphics[width=3.5in]{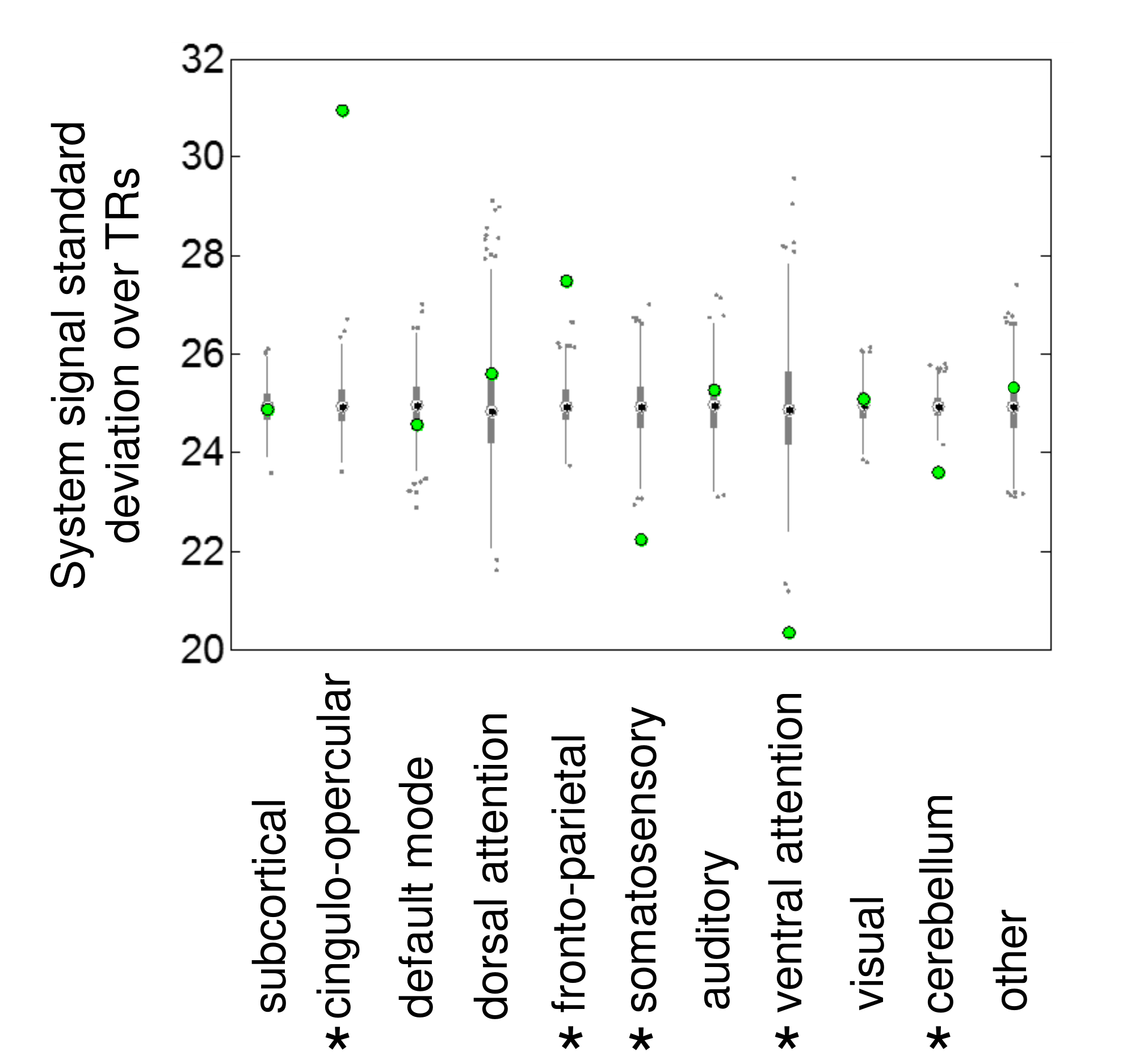}} 
	\caption{\textbf{System-level variability across TRs.} Plot of system-level variability in BOLD time series across TRs. Boxes and whiskers illustrate the distribution of system average standard deviation of BOLD time series over TRs for each system. Green circles illustrate the observed average standard deviations across TRs in each system. Asterisks denote systems in which the observed means were higher or lower than 97.5\% of permuted values.}\label{fig:systemvariability}
\end{figure}

\subsection*{System flexibility is associated with signal alignment}
In the main manuscript, we focus on the trial-to-trial signal alignment with anatomy across the brain as a basic spatiotemporal unity contributing to brain signal organization. We remark that previous analyses have focused on a notion of ``flexibility'' in dynamic networks. This notion is represented by creating a three-dimensional tensor from consecutive or overlapping association matrices (e.g., using correlation, coherence, or mutual information) and observing the changes in node community allegiance over time. Flexible nodes are those that exhibit a relatively high degree of changes in community assignments. Distinct from the current analysis, flexibility occurs at a longer temporal scale since tensors are constructed from many adjacency matrices over time, which in turn are computed over many TRs of time series data. In this context, it is interesting to consider whether the anatomical alignment of BOLD signals at the TR level is associated with the expression of flexibility at the temporal network level.

To explore the relationship between flexible systems and signal alignment with anatomy, we explicitly computed flexibility and associated these values within each participant's alignment values. Specifically, we followed \cite{bassett2013robust} to examine the dynamic community structure of the current data. We constructed a multilayer temporal network for each subject by computing the correlation between all region pairs. Correlations whose significance did not pass false discovery rate correction with a threshold of 0.05 were set to zero \cite{bassett2011dynamic} for each of 10 non-overlapping windows (40 TRs apiece) over the entire task session. We maximized multilayer modularity $Q_{ml}$ following \cite{Mucha2010}:

$$Q_{ml} = \frac{1}{2\mu}\sum_{ijlr}^{}\bigg\{\bigg(A_{ijl}-\gamma_{l}\frac{k_{il}k_{jl}}{2m_{l}}\bigg)\delta_{lr}+\delta_{ij}C_{jlr}\bigg\}\delta(g_{il},g_{jr}),$$

\noindent where the adjacency matrix of layer $l$ (i.e., time window number $l$) has components $A_{ill}$, $\gamma_{l}$ is the resolution parameter of layer $l$, $g_{il}$ gives the community assignment of node $i$ in layer $l$, $g_{jr}$ gives the community assignment of node $j$ in layer $r$, $C_{jlr}$ is the connection strength between node $j$ in layer $r$ and node $j$ in layer $l$ (see the discussion below), $k_{il}$ is the strength of node $i$ in layer $l$, $2\mu = \sum_{jr}^{}k_{jr}, k_{jl} = k_{jl} +c_{jl}$, and $c_{jl} = \sum_{r}^{}C_{jlr}$. For simplicity, we set the resolution parameter $\gamma_{l}$ to unity and we have set all non-zero $C_{jlr}$ to a constant $C$, which we term the ‘inter-layer coupling’. Here, we compute community assignments for $C = 1$. For each subject, we optimized the quality function 100 times and identified a representative consensus partition for all nodes over all windows \cite{bassett2013multiscale}.

Within each subject's consensus partition, we defined the flexibility of a node $f_{i}$ to be the number of times that node changed modular assignment throughout the session, normalized by the total number of changes that were possible (i.e., by the number of consecutive pairs of layers in the multilayer framework) \cite{bassett2011dynamic}. We then defined the flexibility of each system examined in the main text as the mean flexibility over all nodes in the subsystem. Finally, we calculated the Pearson correlation coefficient between the mean system flexibility and mean system alignment. Notably, we found that mean flexibility and mean alignment were significantly positively correlated with one another across systems ($R = 0.15, p = 0.015)$. This result indicates that trial-level signal alignment across the brain is positively associated with the expression of temporal system flexibility over much larger temporal scales across brain networks.

\subsection*{Signal alignment in anatomy is similar to function}

To provide further intuition for our observation that signal alignment is greatest in the most flexible systems, we highlight that the seemingly counterintuitive observation -- the regions whose signals are aligned with the underlying structural network are also the regions that display the greatest network flexibility across time points -- is mathematically justifiable. The reason comes from the close relationship between the graph Fourier transform (GFT) and Principal Component Analysis (PCA) \cite{leonardi2013principal, viviani2005functional}, when the network represents pairwise correlations between brain regions across time points. In this case, from PCA, the signal distribution that explains most of the variation of signals across time points is represented by the eigenvectors associated with large eigenvalues of the correlation matrix (functional connectivity network). Simultaneously, recall from GFT that the eigenvectors associated with large eigenvalues of the correlation matrix depict signal distributions that align well with the network \cite{sandryhaila2013discrete,ma2015}. This implies that the signal distribution that aligns with the network representing pairwise correlations between brain regions also varies the most across time points. Stated simply, it can be derived mathematically that the most flexible systems are most aligned with the functional network constructed from those signals \cite{huang2016graph}. 

Thus, crucial to the current result, if the correlation matrix's eigenspectrum is similar to the anatomical network, the most anatomically aligned functional signals will also be related to the most functionally aligned and flexible signals. In order to evaluate this relationship empirically, we examined the correlation between signals aligned with the structural networks and signals aligned with the functional networks. For each subject in the experiment, we constructed the subject's correlation matrix across all TRs and performed the same decomposition of each BOLD TR from our primary analysis using each subject's (1) anatomical network and then repeated this analysis using (2) the correlation matrix across all BOLD TRs. Next, we examined the similarity between the signals aligned with the functional correlation network and those aligned with the anatomical network. A significant correlation would indicate that the decomposition of the aligned portion of signals is organized with the functional correlation network similarly to its decomposition into the anatomical network, suggesting that the relationships between flexibility and alignment is expressed both in the subject's anatomical and mean functional correlation network. 

We find a strong correlation between the anatomically and functionally aligned signals ($R = 0.814$, $p = 1.613 \times 10^{-27}$). This empirical relationship indicates that signals most aligned with the functional network are those most aligned with the anatomical network. Thus, the eigenspectrum of the functional network that contributes to the tendency for flexible functional signals to align is related to the anatomical eigenspectrum cross subjects. This highlights a fundamental relationship guiding BOLD signal dynamics in anatomical and functional network analysis. Both anatomical and functional correlation networks demonstrate an eigenspectrum that organizes flexible activity in the human brain.

\newpage
\subsection*{Aligned and liberal signal associations with behavior across task conditions}

In the main text, we focus on the relationship between variability in signal alignment and variability in switch costs due to our emphasis on cognitive control in human brain networks. For reference, here we include a full table of results associating median response times in each condition (no switching, switching, and switch costs) with aligned and liberal signals across the 28 subjects. The results indicate that within the current signal decomposition using anatomical networks, only the liberal signals demonstrate a relationship with behavioral variability and specifically with switch costs.

\begin{table}[ht!]
	\caption{\textbf{Pearson's partial correlation coefficients: all median response time associations with signal liberality and alignment.}} 
	
	\begin{centering}
	\begin{tabular}{c c c c c c c} 
	\hline\hline 
		\noalign{\vskip 2mm} 
	Behavior & Lib Fix & Lib NS & Lib S & Ali Fix & Ali NS & Ali S \\ [0.5ex] 
	\hline 
	\noalign{\vskip 2mm}    
	No Switch     &  0.2223   &   0.3503  &  0.2207    &   0.0328  &   0.1134  &  -0.0907\\
	Switch        &  0.4264   &   0.5563  &  0.4026    &   0.1962  &   0.2298  &   0.0090\\
	Switch Cost   &  0.6181   &   0.7137  &  0.5669    &   0.3912  &   0.3431  &   0.1599\\ [1ex]
	\hline 
	\end{tabular}
	\label{table:nonlin} 
	\end{centering}\\
	Pearson's correlation coefficients using average framewise displacement as a covariate. Rows contain partial correlation coefficients associating each median response time at the subject level with their graph alignment. Columns represent the signal type including the fixation, no switch, and switch block signals. Lib = Liberal, Ali = Aligned, Fix = fixation, NS = no switch, S = switch.
	\end{table}

\begin{table}[ht!]
	\caption{\textbf{Pearson's partial correlation $p$-values: all median response time associations with signal liberality and alignment.}} 
	\begin{centering}
	\begin{tabular}{c c c c c c c} 
	\hline\hline 
		\noalign{\vskip 2mm} 
	Behavior & Lib Fix & Lib NS & Lib S & Ali Fix & Ali NS & Ali S \\ [0.5ex] 
	\hline 
	\noalign{\vskip 2mm}    
	No Switch   &  	  0.2650 &   0.0733    &   0.2686 &   0.8709 &   0.5734 &   0.6526\\
	Switch      &     0.0266 &   0.0026    &   0.0374 &   0.3267 &   0.2488 &   0.9644\\
	Switch Cost &     0.0006 &   $<$0.0001 &   0.0020 &   0.0436 &   0.0798 &   0.4258\\[1ex]
	\hline 
	\end{tabular}
	\label{table:nonlin2} 
	\end{centering}\\
	Rows contain $p$-values corresponding to correlation coefficients in Table 1. Columns represent the signal type including the fixation, no switch, and switch block signals. Lib = Liberal, Ali = Aligned, Fix = fixation, NS = no switch, S = switch.
	\end{table}
	
\subsection*{Inclusion of age and sex as covariates}

In the main text, we relate alignment and liberality to behavior using subjects' average framewise displacement as covariates. To examine whether our results are stable when accounting for subjects' age and sex, we recompute partial correlations for each task condition and alignment and liberality values using average framewise displacement, age, and sex as covariates. Our results remain significant with similar correlation values across conditions when including these covariates. 

\begin{table}[ht!]
	\caption{\textbf{Pearson's partial correlation coefficients: all median response time associations with signal liberality and alignment.}} 
	
	\begin{centering}
		\begin{tabular}{c c c c c c c} 
			\hline\hline 
			\noalign{\vskip 2mm} 
			Behavior & Lib Fix & Lib NS & Lib S & Ali Fix & Ali NS & Ali S \\ [0.5ex] 
			\hline 
			\noalign{\vskip 2mm}    
			No Switch   & 0.3508  &  0.4990  &  0.4116 &   0.1168 &   0.1842 &  -0.0042\\
			Switch 	    & 0.5009  &  0.6443  &  0.5390 &   0.2319 &   0.2390 &   0.0527\\
			Switch Cost & 0.6028  &  0.7052  &  0.5984 &   0.3459 &   0.2629 &   0.1237\\ [1ex]
			\hline 
		\end{tabular}
		\label{table:nonlin} 
	\end{centering}\\
	Pearson's correlation coefficients using average framewise displacement, age, and sex as covariates. Rows contain partial correlation coefficients associating each median response time at the subject level with their graph liberality. Columns represent the signal type including the fixation, no switch, and switch block signals. Lib = Liberal, Ali = Aligned, Fix = fixation, NS = no switch, S = switch.
\end{table}

\begin{table}[ht!]
	\caption{\textbf{Pearson's partial correlation $p$-values: all median response time associations with signal liberality and alignment.}} 
	\begin{centering}
		\begin{tabular}{c c c c c c c} 
			\hline\hline 
			\noalign{\vskip 2mm} 
			Behavior & Lib Fix & Lib NS & Lib S & Ali Fix & Ali NS & Ali S \\ [0.5ex] 
			\hline 
			\noalign{\vskip 2mm}    
			No Switch	 & 0.0856  &  0.0111  &  0.0409  &  0.5783  &  0.3780  &  0.9842\\
			Switch 		 & 0.0107  &  0.0005  &  0.0054  &  0.2646  &  0.2499  &  0.8023\\
			Switch Cost	 & 0.0014  &  0.0001  &  0.0016  &  0.0903  &  0.2042  &  0.5559\\ [1ex]
			\hline 
		\end{tabular}
		\label{table:nonlin} 
	\end{centering}\\
	Rows contain $p$-values corresponding to correlation coefficients in Table 3. Columns represent the signal type including the fixation, no switch, and switch block signals. Lib = Liberal, Ali = Aligned, Fix = fixation, NS = no switch, S = switch.
\end{table}

\subsection*{Robustness of aligned and liberal signals to parameter selection}

In the main text, we report results where we represent the liberal signals $\bbx_\L$ and aligned signals $\bbx_\HH$ using cut points for $K_\L$ and $K_\HH$ of the ten lowest (of $K_\L$) and highest ($K_\HH$) values. To test the robustness of aligned and liberal signals to variations in these parameters, we decompose the observed BOLD signals into $\bbx_\L$ and $\bbx_\HH$ under the choice of $K_\L$ and $K_\HH$ from five below to five above the choice of parameters studied in the main manuscript. We then compute the correlation coefficient among $\bbx_\L$ and $\bbx_\HH$ across all TRs for all subjects to examine the stability of aligned and liberal signals. We find that the observed signal decomposition is stable across parameter choices: for $\bbx_\L$ (liberal signals), the values are correlated at a mean of $R = 0.95$ (St.D. $0.03$); for $\bbx_\HH$ (aligned signals), the values are correlated at a mean of $R = 0.99$ (St.D. $0.02$). Thus, the signals examined in the main text are highly stable over the choice of definition for the range of aligned and liberal signals, and especially for liberal signals.

\subsection*{Null permutation test for the significance of the anatomical organization of signals}

In the current analysis, the aligned signals are found to be the most concentrated in systems including those known to be functionally flexible in prior studies \cite{mattar2016flexible,braun2015dynamic}. It is interesting to examine the complexity of anatomical contributions to these signals. We consider whether anatomically aligned signals are organized by the simple contributions of nearest neighbor nodes in anatomical networks (i.e., only the nodes that share direct connections to each node) \emph{versus} the specific configuration of the entire anatomical network. This allows us to identify whether the BOLD signals inclusive of the most flexible systems in the brain are aligned with anatomy as a function of the contribution of the entire network, which would potentially suggest that functionally flexible systems are organized by weighted contributions from the entire brain.

To test this possibility, we conduct a non-parametric permutation test. Specifically, we randomize each individual's network 2000 times preserving the strength and degree sequence of nodes in the network. Then, we decompose the observed BOLD value for each region into its corresponding node in the randomized network across all TRs. We then calculate over subjects the average variance in original BOLD signals accounted for by the eigenvectors associated with the aligned signals for each random permutation. We summarize these values as the mean variance accounted for across all null permutations and compare it to the mean variance accounted for by the original anatomical networks. We observe that the variance accounted for in the real networks exceeds that accounted for in 100\% of the null permutations. This indicates that the BOLD signals are organized by anatomical features at a greater topological length across the networks than the direct connections among nodes (See Fig.~\ref{fig:shufflenetworks}). This finding indicates that signals that depend on underlying anatomy represent complex contributions from anatomical network topology. 

The behavioral relevance of this anatomically-aligned organization may be established in future studies. In particular, when the underlying network is constructed from fMRI measurements (functional similarity networks constructed from correlations), it is known that the components of BOLD signals that are the most aligned to the network also explain the most variation across time \cite{huang2016graph}. Here also we find that signals aligned with anatomical connectivity contain BOLD measurements that are the most variable over time in cingulo-opercular and fronto-parietal systems. Functional connectivity (e.g., correlation, coherence, and mutual information) networks evince similar but not completely overlapping topology when compared to anatomical networks \cite{bullmore2009complex}. Thus, analyzing the alignment of measured correlations on anatomical networks may provide an important extension to understand signals in the time domain in the human connectome. Mathematically, the same phenomenon for functional as well as anatomical networks implies that the eigenspectrum of the anatomical network is not that different from functional connectivity. Speculatively, this could provide a meaningful and efficient method to reliably recover anatomical networks from functional connectivity networks.

\begin{figure} [h!]
	\centerline{\includegraphics[width=3.5in]{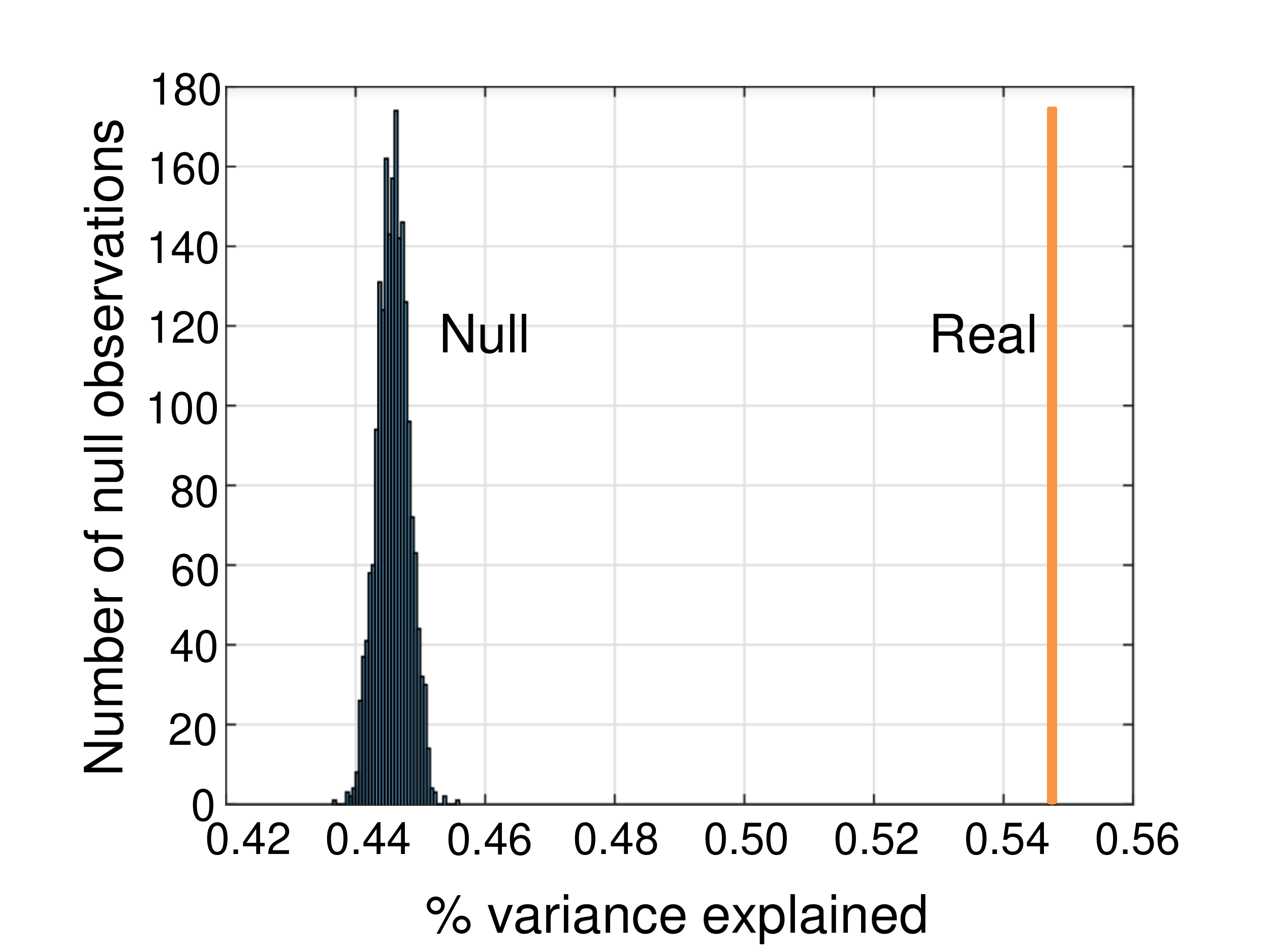}} 
	\caption{\textbf{Aligned signals are organized by topologically distant anatomical organization.} Left: in the null permutations where the edges of the anatomical networks were randomized, a mean of 0.455\% variance is accounted for by the eigenvectors corresponding to the aligned parts of the signal. In the observed data, a mean of 0.547\% of variance is accounted for by the anatomical networks in their true configurations.} \label{fig:shufflenetworks}
\end{figure}

\subsection*{Alternate measures of function and anatomy}
Do anatomy-decomposed functional BOLD signals better related to switch costs than other measures with similar intuitions? To test this possibility, we consider four comparison analyses involving the subcortical system from which the behaviorally relevant signals in our primary analyses originated. First, we compute the mean BOLD signal variance over all subcortical regions for each TR. This correlation represents a measure of differences in BOLD signals at each time measurement without considering its anatomical network context. We then relate the average BOLD signal variance over TRs to switch costs and find no significant relationship ($R = -0.09$, $p = 0.68$). Second, we compute the mean node \emph{strength} -- the sum of connections to the node -- for each region in the subcortical system. This represents the overall influence of the subcortical regions in the broader anatomical network. We calculate the Pearson correlation coefficient between the mean subcortical node strength and switch costs across subjects, and we find no significant relationship ($R = -0.15$, $p = 0.48$). Finally, within each subject, we compute the mean Pearson correlation coefficient between each BOLD signal TR and node strengths within the subcortical system, then we compute the Pearson correlation coefficient between these values and switch costs across subjects, finding no significant relationship ($R = 0.17$, $p = 0.41$). Taken together, these results demonstrate that simple measures of BOLD variation, local anatomical network influences, and the relationship between the two are insufficient to account for switch cost variability over subjects. Indeed, decomposing signals in the context of the entire anatomical network provides superior associative value for the cognitive control measure. 

\begin{figure} [h!]
	\centerline{\includegraphics[width=7in]{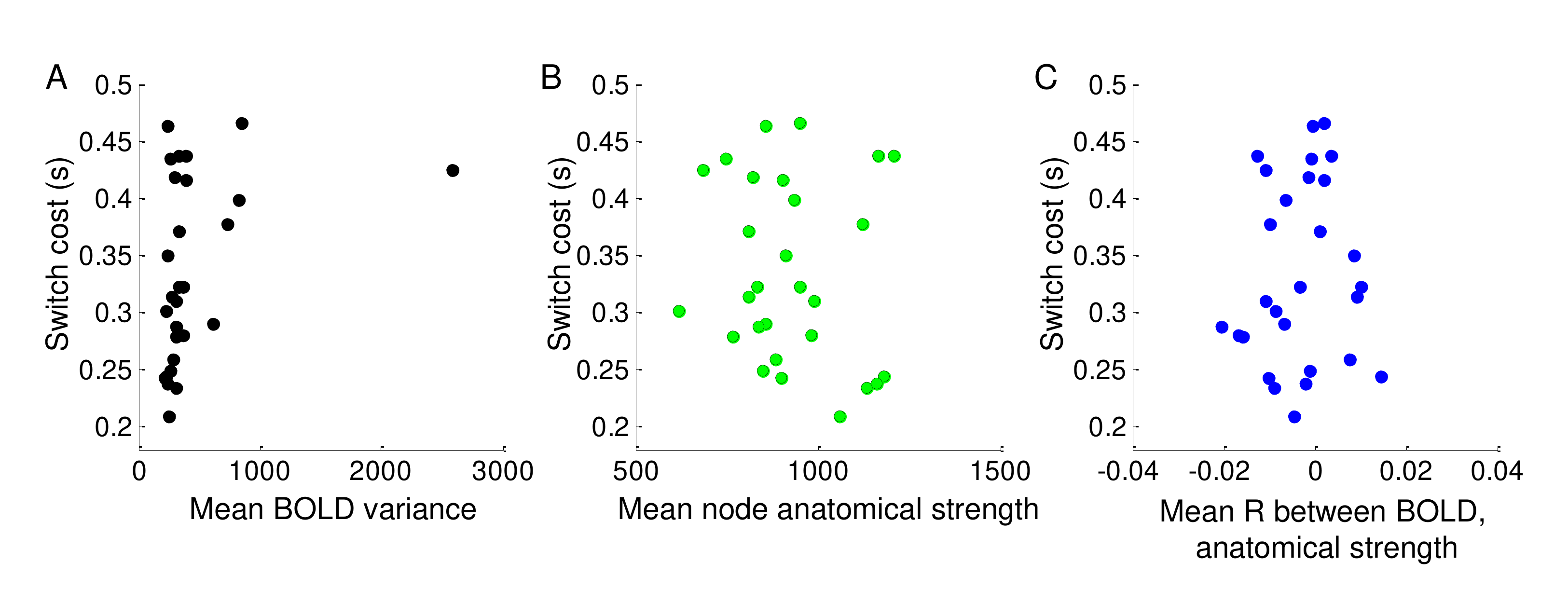}} 
	\caption{\textbf{Alternate measures of framewise BOLD variance, anatomy, and function-anatomy relationships in subcortical systems do not associate with switch costs.} \emph{(A)} The relationship between mean BOLD variance in the subcortical system and switch costs. \emph{(B)} The relationship between mean anatomical node degree in the subcortical system and switch costs. \emph{(C)} The relationship between the correlation between BOLD signals and anatomical node degree across regions in the subcortical systems and switch costs.} \label{fig:shufflenetworks}
\end{figure}

\newpage
\subsection*{Anatomical network null model for behavioral correlations with switch costs.} 
\begin{figure} [h!]
\centerline{\includegraphics[width=3.5in]{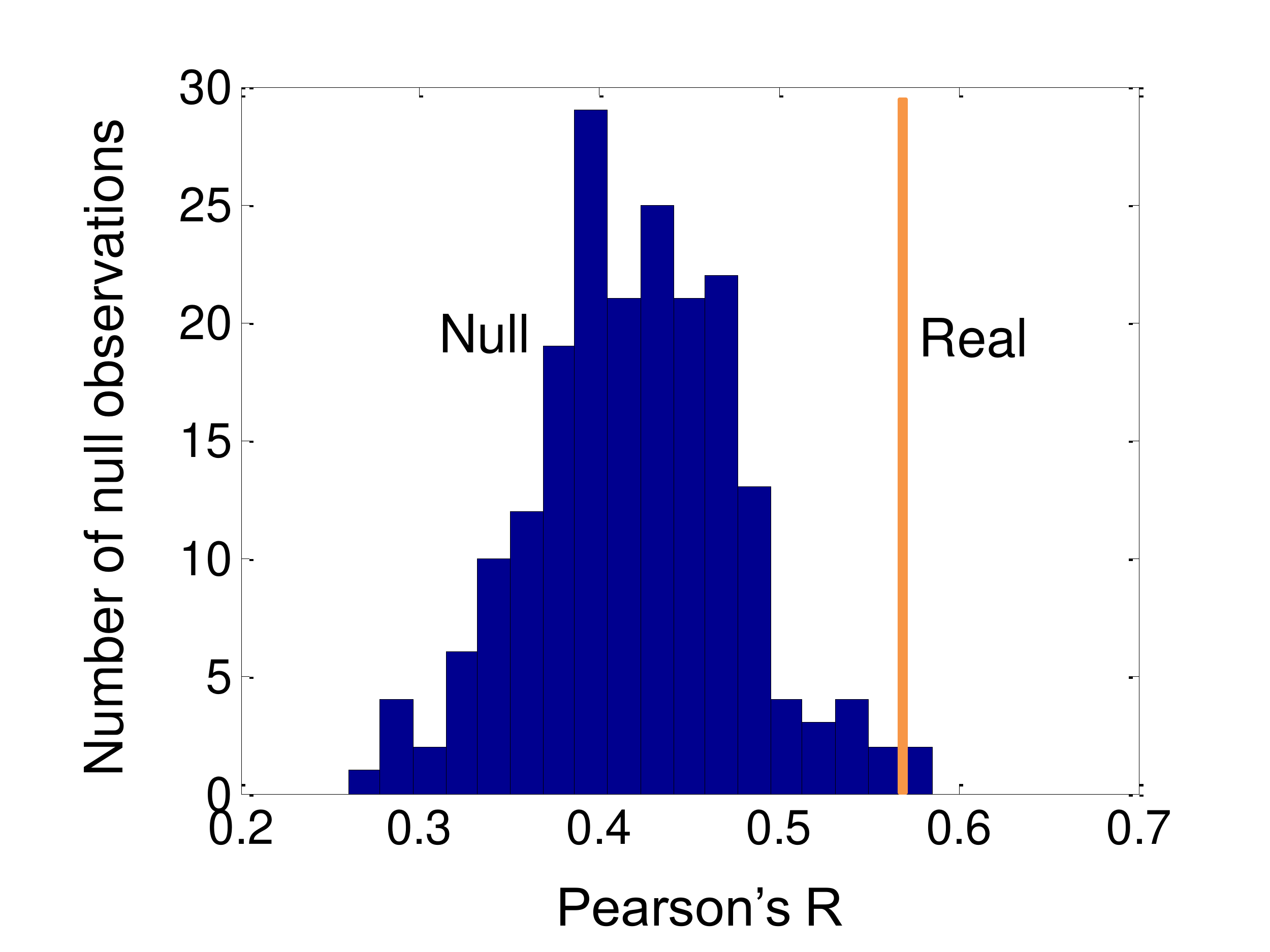}} 
	\caption{\textbf{A null permutation test demonstrates that specific anatomical network organization is required to identify behaviorally relevant BOLD signals.} Blue bars in the histogram illustrate the Pearson's correlation coefficients between switch costs and liberal signals over null permutations. Orange line designates the observed correlation value of the real network.} \label{fig:shufflenetworkscorrelation}
\end{figure}

To examine the importance of the specific anatomical configuration of brain networks for the association between liberal signals and switch costs, we performed a null permutation test. Specifically, we performed 200 permutations in which we generated a random network preserving the strength and degree distributions of the original anatomical networks for each subject. Then, we repeated the BOLD signal decomposition into the null anatomical network for each subject and correlated the liberal signals with switch costs in each permutation to generate a null distribution of correlations. From this null distribution, we computed the proportion of null correlation values greater than the observed correlation when using the real anatomical networks. We find that the observed correlation is greater than 99\% of correlations (Fig.~\ref{fig:shufflenetworkscorrelation}). This result indicates that the specific configuration of the true anatomical network drives the correlation between liberal signal alignment and switch costs.

\subsection*{Relationships between motion and signal alignment.} 
In the results reported in the main text and supplement, we included a motion covariate (average framewise displacement across BOLD TRs) in all analyses. Motion is not significantly correlated with switch cost ($R = 0.26$, $p = 0.16$), but is correlated significantly with liberal signals ($R = 0.77$, $p = 1.10 \times 10^{-6}$), justifying its inclusion as a covariate in behavioral analyses. Importantly, when motion is not included as a covariate in the partial correlation between liberal signals and switch costs, the correlation ($R = 0.59$ , $ p = 0.001$) is highly similar to that observed when including motion as a covariate ($R = 0.57$, $p = 0.002$). This finding indicates that the behaviorally relevant portion of the liberal signals is not driven by motion.


\bibliographystyle{naturemag}
\bibliography{JDMReferences}

\begin{thebibliography}{10}
\expandafter\ifx\csname url\endcsname\relax
  \def\url#1{\texttt{#1}}\fi
\expandafter\ifx\csname urlprefix\endcsname\relax\def\urlprefix{URL }\fi
\providecommand{\bibinfo}[2]{#2}
\providecommand{\eprint}[2][]{\url{#2}}

\bibitem{rogers1995costs}
\bibinfo{author}{Rogers, R.~D.} \& \bibinfo{author}{Monsell, S.}
\newblock \bibinfo{title}{Costs of a predictible switch between simple
  cognitive tasks.}
\newblock \emph{\bibinfo{journal}{Journal of experimental psychology: General}}
  \textbf{\bibinfo{volume}{124}}, \bibinfo{pages}{207} (\bibinfo{year}{1995}).

\bibitem{szczepanski2014insights}
\bibinfo{author}{Szczepanski, S.~M.} \& \bibinfo{author}{Knight, R.~T.}
\newblock \bibinfo{title}{Insights into human behavior from lesions to the
  prefrontal cortex}.
\newblock \emph{\bibinfo{journal}{Neuron}} \textbf{\bibinfo{volume}{83}},
  \bibinfo{pages}{1002--1018} (\bibinfo{year}{2014}).

\bibitem{clark2012specific}
\bibinfo{author}{Clark, L.~R.} \emph{et~al.}
\newblock \bibinfo{title}{Specific measures of executive function predict
  cognitive decline in older adults}.
\newblock \emph{\bibinfo{journal}{Journal of the International
  Neuropsychological Society}} \textbf{\bibinfo{volume}{18}},
  \bibinfo{pages}{118--127} (\bibinfo{year}{2012}).

\bibitem{richland2013early}
\bibinfo{author}{Richland, L.~E.} \& \bibinfo{author}{Burchinal, M.~R.}
\newblock \bibinfo{title}{Early executive function predicts reasoning
  development}.
\newblock \emph{\bibinfo{journal}{Psychological science}}
  \textbf{\bibinfo{volume}{24}}, \bibinfo{pages}{87--92}
  (\bibinfo{year}{2013}).

\bibitem{davis2010independent}
\bibinfo{author}{Davis, J.~C.}, \bibinfo{author}{Marra, C.~A.},
  \bibinfo{author}{Najafzadeh, M.} \& \bibinfo{author}{Liu-Ambrose, T.}
\newblock \bibinfo{title}{The independent contribution of executive functions
  to health related quality of life in older women}.
\newblock \emph{\bibinfo{journal}{BMC geriatrics}}
  \textbf{\bibinfo{volume}{10}}, \bibinfo{pages}{16} (\bibinfo{year}{2010}).

\bibitem{gunaydin2016cortico}
\bibinfo{author}{Gunaydin, L.~A.} \& \bibinfo{author}{Kreitzer, A.~C.}
\newblock \bibinfo{title}{Cortico-basal ganglia circuit function in psychiatric
  disease}.
\newblock \emph{\bibinfo{journal}{Annual review of physiology}}
  \textbf{\bibinfo{volume}{78}}, \bibinfo{pages}{327--350}
  (\bibinfo{year}{2016}).

\bibitem{casey2004early}
\bibinfo{author}{Casey, B.} \emph{et~al.}
\newblock \bibinfo{title}{Early development of subcortical regions involved in
  non-cued attention switching}.
\newblock \emph{\bibinfo{journal}{Developmental science}}
  \textbf{\bibinfo{volume}{7}}, \bibinfo{pages}{534--542}
  (\bibinfo{year}{2004}).

\bibitem{cole2013multi}
\bibinfo{author}{Cole, M.~W.} \emph{et~al.}
\newblock \bibinfo{title}{Multi-task connectivity reveals flexible hubs for
  adaptive task control}.
\newblock \emph{\bibinfo{journal}{Nature neuroscience}}
  \textbf{\bibinfo{volume}{16}}, \bibinfo{pages}{1348--1355}
  (\bibinfo{year}{2013}).

\bibitem{heyder2004cortico}
\bibinfo{author}{Heyder, K.}, \bibinfo{author}{Suchan, B.} \&
  \bibinfo{author}{Daum, I.}
\newblock \bibinfo{title}{Cortico-subcortical contributions to executive
  control}.
\newblock \emph{\bibinfo{journal}{Acta psychologica}}
  \textbf{\bibinfo{volume}{115}}, \bibinfo{pages}{271--289}
  (\bibinfo{year}{2004}).

\bibitem{luk2012cognitive}
\bibinfo{author}{Luk, G.}, \bibinfo{author}{Green, D.~W.},
  \bibinfo{author}{Abutalebi, J.} \& \bibinfo{author}{Grady, C.}
\newblock \bibinfo{title}{Cognitive control for language switching in
  bilinguals: A quantitative meta-analysis of functional neuroimaging studies}.
\newblock \emph{\bibinfo{journal}{Language and cognitive processes}}
  \textbf{\bibinfo{volume}{27}}, \bibinfo{pages}{1479--1488}
  (\bibinfo{year}{2012}).

\bibitem{hikosaka2010switching}
\bibinfo{author}{Hikosaka, O.} \& \bibinfo{author}{Isoda, M.}
\newblock \bibinfo{title}{Switching from automatic to controlled behavior:
  cortico-basal ganglia mechanisms}.
\newblock \emph{\bibinfo{journal}{Trends in cognitive sciences}}
  \textbf{\bibinfo{volume}{14}}, \bibinfo{pages}{154--161}
  (\bibinfo{year}{2010}).

\bibitem{hosoda2012neural}
\bibinfo{author}{Hosoda, C.}, \bibinfo{author}{Hanakawa, T.},
  \bibinfo{author}{Nariai, T.}, \bibinfo{author}{Ohno, K.} \&
  \bibinfo{author}{Honda, M.}
\newblock \bibinfo{title}{Neural mechanisms of language switch}.
\newblock \emph{\bibinfo{journal}{Journal of Neurolinguistics}}
  \textbf{\bibinfo{volume}{25}}, \bibinfo{pages}{44--61}
  (\bibinfo{year}{2012}).

\bibitem{leunissen2014subcortical}
\bibinfo{author}{Leunissen, I.} \emph{et~al.}
\newblock \bibinfo{title}{Subcortical volume analysis in traumatic brain
  injury: the importance of the fronto-striato-thalamic circuit in task
  switching}.
\newblock \emph{\bibinfo{journal}{Cortex}} \textbf{\bibinfo{volume}{51}},
  \bibinfo{pages}{67--81} (\bibinfo{year}{2014}).

\bibitem{yehene2008basal}
\bibinfo{author}{Yehene, E.}, \bibinfo{author}{Meiran, N.} \&
  \bibinfo{author}{Soroker, N.}
\newblock \bibinfo{title}{Basal ganglia play a unique role in task switching
  within the frontal-subcortical circuits: evidence from patients with focal
  lesions}.
\newblock \emph{\bibinfo{journal}{Journal of Cognitive Neuroscience}}
  \textbf{\bibinfo{volume}{20}}, \bibinfo{pages}{1079--1093}
  (\bibinfo{year}{2008}).

\bibitem{sporns2005}
\bibinfo{author}{Sporns, O.}, \bibinfo{author}{Tononi, G.} \&
  \bibinfo{author}{K\"{o}tter, R.}
\newblock \bibinfo{title}{The human connectome: A structural description of the
  human brain}.
\newblock \emph{\bibinfo{journal}{PLoS Computational Biology}}
  \textbf{\bibinfo{volume}{1}}, \bibinfo{pages}{e42} (\bibinfo{year}{2005}).

\bibitem{alstott2009}
\bibinfo{author}{Alstott, J.}, \bibinfo{author}{Breakspear, M.},
  \bibinfo{author}{Hagmann, P.}, \bibinfo{author}{Cammoun, L.} \&
  \bibinfo{author}{Sporns, O.}
\newblock \bibinfo{title}{{Modeling the impact of lesions in the human brain.}}
\newblock \emph{\bibinfo{journal}{PLoS computational biology}}
  \textbf{\bibinfo{volume}{5}}, \bibinfo{pages}{e1000408}
  (\bibinfo{year}{2009}).

\bibitem{Hermundstad2013}
\bibinfo{author}{Hermundstad, A.~M.} \emph{et~al.}
\newblock \bibinfo{title}{Structural foundations of resting-state and
  task-based functional connectivity in the human brain}.
\newblock \emph{\bibinfo{journal}{Proceedings of the National Academy of
  Sciences}} \textbf{\bibinfo{volume}{110}}, \bibinfo{pages}{6169--6174}
  (\bibinfo{year}{2013}).

\bibitem{Honey2007}
\bibinfo{author}{Honey, C.~J.}, \bibinfo{author}{K\"{o}tter, R.},
  \bibinfo{author}{Breakspear, M.} \& \bibinfo{author}{Sporns, O.}
\newblock \bibinfo{title}{Network structure of cerebral cortex shapes
  functional connectivity on multiple time scales}.
\newblock \emph{\bibinfo{journal}{Proceedings of the National Academy of
  Sciences}} \textbf{\bibinfo{volume}{104}}, \bibinfo{pages}{10240--10245}
  (\bibinfo{year}{2007}).

\bibitem{medaglia2015cognitive}
\bibinfo{author}{Medaglia, J.~D.}, \bibinfo{author}{Lynall, M.-E.} \&
  \bibinfo{author}{Bassett, D.~S.}
\newblock \bibinfo{title}{Cognitive network neuroscience}.
\newblock \emph{\bibinfo{journal}{Journal of cognitive neuroscience}}
  (\bibinfo{year}{2015}).

\bibitem{sporns2014challenges}
\bibinfo{author}{Sporns, O.}
\newblock \bibinfo{title}{Contributions and challenges for network models in
  cognitive neuroscience}.
\newblock \emph{\bibinfo{journal}{Nature Neuroscience}}
  \textbf{\bibinfo{volume}{17}}, \bibinfo{pages}{652--660}
  (\bibinfo{year}{2014}).

\bibitem{power2011functional}
\bibinfo{author}{Power, J.~D.} \emph{et~al.}
\newblock \bibinfo{title}{Functional network organization of the human brain}.
\newblock \emph{\bibinfo{journal}{Neuron}} \textbf{\bibinfo{volume}{72}},
  \bibinfo{pages}{665--678} (\bibinfo{year}{2011}).

\bibitem{navon1977forest}
\bibinfo{author}{Navon, D.}
\newblock \bibinfo{title}{Forest before trees: The precedence of global
  features in visual perception}.
\newblock \emph{\bibinfo{journal}{Cognitive psychology}}
  \textbf{\bibinfo{volume}{9}}, \bibinfo{pages}{353--383}
  (\bibinfo{year}{1977}).

\bibitem{cammoun2012}
\bibinfo{author}{Cammoun, L.} \emph{et~al.}
\newblock \bibinfo{title}{Mapping the human connectome at multiple scales with
  diffusion spectrum mri}.
\newblock \emph{\bibinfo{journal}{Journal of Neuroscience Methods}}
  \textbf{\bibinfo{volume}{203}}, \bibinfo{pages}{386--397}
  (\bibinfo{year}{2012}).

\bibitem{diedrichsen2009probabilistic}
\bibinfo{author}{Diedrichsen, J.}, \bibinfo{author}{Balsters, J.~H.},
  \bibinfo{author}{Flavell, J.}, \bibinfo{author}{Cussans, E.} \&
  \bibinfo{author}{Ramnani, N.}
\newblock \bibinfo{title}{A probabilistic mr atlas of the human cerebellum}.
\newblock \emph{\bibinfo{journal}{Neuroimage}} \textbf{\bibinfo{volume}{46}},
  \bibinfo{pages}{39--46} (\bibinfo{year}{2009}).

\bibitem{Gu2015}
\bibinfo{author}{Gu, S.} \emph{et~al.}
\newblock \bibinfo{title}{Controllability of structural brain networks}.
\newblock \emph{\bibinfo{journal}{Nature Communications}}
  \textbf{\bibinfo{volume}{6}}, \bibinfo{pages}{8414} (\bibinfo{year}{2015}).

\bibitem{braun2015dynamic}
\bibinfo{author}{Braun, U.} \emph{et~al.}
\newblock \bibinfo{title}{Dynamic reconfiguration of frontal brain networks
  during executive cognition in humans}.
\newblock \emph{\bibinfo{journal}{Proceedings of the National Academy of
  Sciences}} \textbf{\bibinfo{volume}{112}}, \bibinfo{pages}{11678--11683}
  (\bibinfo{year}{2015}).

\bibitem{mayhew2016global}
\bibinfo{author}{Mayhew, S.~D.} \emph{et~al.}
\newblock \bibinfo{title}{Global signal modulation of single-trial fmri
  response variability: Effect on positive vs negative bold response
  relationship}.
\newblock \emph{\bibinfo{journal}{NeuroImage}} \textbf{\bibinfo{volume}{133}},
  \bibinfo{pages}{62--74} (\bibinfo{year}{2016}).

\bibitem{sekutowicz2016striatal}
\bibinfo{author}{Sekutowicz, M.} \emph{et~al.}
\newblock \bibinfo{title}{Striatal activation as a neural link between
  cognitive and perceptual flexibility}.
\newblock \emph{\bibinfo{journal}{NeuroImage}} \textbf{\bibinfo{volume}{141}},
  \bibinfo{pages}{393--398} (\bibinfo{year}{2016}).

\bibitem{liston2006anterior}
\bibinfo{author}{Liston, C.}, \bibinfo{author}{Matalon, S.},
  \bibinfo{author}{Hare, T.~A.}, \bibinfo{author}{Davidson, M.~C.} \&
  \bibinfo{author}{Casey, B.}
\newblock \bibinfo{title}{Anterior cingulate and posterior parietal cortices
  are sensitive to dissociable forms of conflict in a task-switching paradigm}.
\newblock \emph{\bibinfo{journal}{Neuron}} \textbf{\bibinfo{volume}{50}},
  \bibinfo{pages}{643--653} (\bibinfo{year}{2006}).

\bibitem{cole2014intrinsic}
\bibinfo{author}{Cole, M.~W.}, \bibinfo{author}{Bassett, D.~S.},
  \bibinfo{author}{Power, J.~D.}, \bibinfo{author}{Braver, T.~S.} \&
  \bibinfo{author}{Petersen, S.~E.}
\newblock \bibinfo{title}{Intrinsic and task-evoked network architectures of
  the human brain}.
\newblock \emph{\bibinfo{journal}{Neuron}} \textbf{\bibinfo{volume}{83}},
  \bibinfo{pages}{238--251} (\bibinfo{year}{2014}).

\bibitem{watanabe2014energy}
\bibinfo{author}{Watanabe, T.}, \bibinfo{author}{Masuda, N.},
  \bibinfo{author}{Megumi, F.}, \bibinfo{author}{Kanai, R.} \&
  \bibinfo{author}{Rees, G.}
\newblock \bibinfo{title}{Energy landscape and dynamics of brain activity
  during human bistable perception}.
\newblock \emph{\bibinfo{journal}{Nature communications}}
  \textbf{\bibinfo{volume}{5}} (\bibinfo{year}{2014}).

\bibitem{mattar2016flexible}
\bibinfo{author}{Mattar, M.~G.}, \bibinfo{author}{Betzel, R.~F.} \&
  \bibinfo{author}{Bassett, D.~S.}
\newblock \bibinfo{title}{The flexible brain}.
\newblock \emph{\bibinfo{journal}{Brain}} \textbf{\bibinfo{volume}{139}},
  \bibinfo{pages}{2110--2112} (\bibinfo{year}{2016}).

\bibitem{Miyake2000}
\bibinfo{author}{Miyake, A.}, \bibinfo{author}{Friedman, N.~P.},
  \bibinfo{author}{Emerson, M.~J.}, \bibinfo{author}{Witzki, A.~H.} \&
  \bibinfo{author}{Howerter, A.}
\newblock \bibinfo{title}{The unity and diversity of executive functions and
  their contributions to complex "frontal lobe" tasks: A latent variable
  analysis}.
\newblock \emph{\bibinfo{journal}{Cognitive Psychology}}
  \textbf{\bibinfo{volume}{41}}, \bibinfo{pages}{49--100}
  (\bibinfo{year}{2000}).

\bibitem{fedorenko2014role}
\bibinfo{author}{Fedorenko, E.}
\newblock \bibinfo{title}{The role of domain-general cognitive control in
  language comprehension}.
\newblock \emph{\bibinfo{journal}{Frontiers in psychology}}
  \textbf{\bibinfo{volume}{5}}, \bibinfo{pages}{335} (\bibinfo{year}{2014}).

\bibitem{hermundstad2014structurally}
\bibinfo{author}{Hermundstad, A.~M.} \emph{et~al.}
\newblock \bibinfo{title}{Structurally-constrained relationships between
  cognitive states in the human brain}.
\newblock \emph{\bibinfo{journal}{PLoS Comput Biol}}
  \textbf{\bibinfo{volume}{10}}, \bibinfo{pages}{e1003591}
  (\bibinfo{year}{2014}).

\bibitem{belleville2008task}
\bibinfo{author}{Belleville, S.}, \bibinfo{author}{Bherer, L.},
  \bibinfo{author}{Lepage, {\'E}.}, \bibinfo{author}{Chertkow, H.} \&
  \bibinfo{author}{Gauthier, S.}
\newblock \bibinfo{title}{Task switching capacities in persons with alzheimer's
  disease and mild cognitive impairment}.
\newblock \emph{\bibinfo{journal}{Neuropsychologia}}
  \textbf{\bibinfo{volume}{46}}, \bibinfo{pages}{2225--2233}
  (\bibinfo{year}{2008}).

\bibitem{kehagia2010neuropsychological}
\bibinfo{author}{Kehagia, A.~A.}, \bibinfo{author}{Barker, R.~A.} \&
  \bibinfo{author}{Robbins, T.~W.}
\newblock \bibinfo{title}{Neuropsychological and clinical heterogeneity of
  cognitive impairment and dementia in patients with parkinson's disease}.
\newblock \emph{\bibinfo{journal}{The Lancet Neurology}}
  \textbf{\bibinfo{volume}{9}}, \bibinfo{pages}{1200--1213}
  (\bibinfo{year}{2010}).

\bibitem{kinnunen2010white}
\bibinfo{author}{Kinnunen, K.~M.} \emph{et~al.}
\newblock \bibinfo{title}{White matter damage and cognitive impairment after
  traumatic brain injury}.
\newblock \emph{\bibinfo{journal}{Brain}} \bibinfo{pages}{awq347}
  (\bibinfo{year}{2010}).

\bibitem{betzel2016optimally}
\bibinfo{author}{Betzel, R.~F.}, \bibinfo{author}{Gu, S.},
  \bibinfo{author}{Medaglia, J.~D.}, \bibinfo{author}{Pasqualetti, F.} \&
  \bibinfo{author}{Bassett, D.~S.}
\newblock \bibinfo{title}{Optimally controlling the human connectome: the role
  of network topology}.
\newblock \emph{\bibinfo{journal}{Scientific Reports}}  (\bibinfo{year}{2016}).

\bibitem{yeh2011estimation}
\bibinfo{author}{Yeh, F.-C.}, \bibinfo{author}{Wedeen, V.~J.} \&
  \bibinfo{author}{Tseng, W.-Y.~I.}
\newblock \bibinfo{title}{Estimation of fiber orientation and spin density
  distribution by diffusion deconvolution}.
\newblock \emph{\bibinfo{journal}{Neuroimage}} \textbf{\bibinfo{volume}{55}},
  \bibinfo{pages}{1054--1062} (\bibinfo{year}{2011}).

\bibitem{fischl2012freesurfer}
\bibinfo{author}{Fischl, B.}
\newblock \bibinfo{title}{Freesurfer}.
\newblock \emph{\bibinfo{journal}{Neuroimage}} \textbf{\bibinfo{volume}{62}},
  \bibinfo{pages}{774--781} (\bibinfo{year}{2012}).

\bibitem{cammoun2012mapping}
\bibinfo{author}{Cammoun, L.} \emph{et~al.}
\newblock \bibinfo{title}{Mapping the human connectome at multiple scales with
  diffusion spectrum mri}.
\newblock \emph{\bibinfo{journal}{Journal of neuroscience methods}}
  \textbf{\bibinfo{volume}{203}}, \bibinfo{pages}{386--397}
  (\bibinfo{year}{2012}).

\bibitem{cieslak2014local}
\bibinfo{author}{Cieslak, M.} \& \bibinfo{author}{Grafton, S.}
\newblock \bibinfo{title}{Local termination pattern analysis: a tool for
  comparing white matter morphology}.
\newblock \emph{\bibinfo{journal}{Brain imaging and behavior}}
  \textbf{\bibinfo{volume}{8}}, \bibinfo{pages}{292--299}
  (\bibinfo{year}{2014}).

\bibitem{hagmann2008mapping}
\bibinfo{author}{Hagmann, P.} \emph{et~al.}
\newblock \bibinfo{title}{Mapping the structural core of human cerebral
  cortex}.
\newblock \emph{\bibinfo{journal}{PLoS Biol}} \textbf{\bibinfo{volume}{6}},
  \bibinfo{pages}{e159} (\bibinfo{year}{2008}).

\bibitem{voogd1998anatomy}
\bibinfo{author}{Voogd, J.} \& \bibinfo{author}{Glickstein, M.}
\newblock \bibinfo{title}{The anatomy of the cerebellum}.
\newblock \emph{\bibinfo{journal}{Trends in cognitive sciences}}
  \textbf{\bibinfo{volume}{2}}, \bibinfo{pages}{307--313}
  (\bibinfo{year}{1998}).

\bibitem{jenkinson2012fsl}
\bibinfo{author}{Jenkinson, M.}, \bibinfo{author}{Beckmann, C.~F.},
  \bibinfo{author}{Behrens, T.~E.}, \bibinfo{author}{Woolrich, M.~W.} \&
  \bibinfo{author}{Smith, S.~M.}
\newblock \bibinfo{title}{Fsl}.
\newblock \emph{\bibinfo{journal}{Neuroimage}} \textbf{\bibinfo{volume}{62}},
  \bibinfo{pages}{782--790} (\bibinfo{year}{2012}).

\bibitem{greve2009accurate}
\bibinfo{author}{Greve, D.~N.} \& \bibinfo{author}{Fischl, B.}
\newblock \bibinfo{title}{Accurate and robust brain image alignment using
  boundary-based registration}.
\newblock \emph{\bibinfo{journal}{Neuroimage}} \textbf{\bibinfo{volume}{48}},
  \bibinfo{pages}{63--72} (\bibinfo{year}{2009}).

\bibitem{zhang2001segmentation}
\bibinfo{author}{Zhang, Y.}, \bibinfo{author}{Brady, M.} \&
  \bibinfo{author}{Smith, S.}
\newblock \bibinfo{title}{Segmentation of brain mr images through a hidden
  markov random field model and the expectation-maximization algorithm}.
\newblock \emph{\bibinfo{journal}{IEEE transactions on medical imaging}}
  \textbf{\bibinfo{volume}{20}}, \bibinfo{pages}{45--57}
  (\bibinfo{year}{2001}).

\bibitem{Jenkinson2002}
\bibinfo{author}{Jenkinson, M.}, \bibinfo{author}{Bannister, P.},
  \bibinfo{author}{Brady, M.} \& \bibinfo{author}{Smith, S.}
\newblock \bibinfo{title}{Improved optimization for the robust and accurate
  linear registration and motion correction of brain images}.
\newblock \emph{\bibinfo{journal}{Neuroimage}} \textbf{\bibinfo{volume}{17}},
  \bibinfo{pages}{825--841} (\bibinfo{year}{2002}).

\bibitem{Chung97}
\bibinfo{author}{Chung, F.}
\newblock \emph{\bibinfo{title}{Spectral graph theory}},
  vol.~\bibinfo{volume}{92} (\bibinfo{publisher}{American Mathematical Soc.},
  \bibinfo{year}{1997}).

\bibitem{sandryhaila2013discrete}
\bibinfo{author}{Sandryhaila, A.} \& \bibinfo{author}{Moura, J.~M.}
\newblock \bibinfo{title}{Discrete signal processing on graphs}.
\newblock \emph{\bibinfo{journal}{IEEE transactions on signal processing}}
  \textbf{\bibinfo{volume}{61}}, \bibinfo{pages}{1644--1656}
  (\bibinfo{year}{2013}).

\bibitem{shuman2013}
\bibinfo{author}{Shuman, D.~I.}, \bibinfo{author}{Narang, S.~K.},
  \bibinfo{author}{Frossard, P.}, \bibinfo{author}{Ortega, A.} \&
  \bibinfo{author}{Vandergheynst, P.}
\newblock \bibinfo{title}{The emerging field of signal processing on graphs:
  Extending high-dimensional data analysis to networks and other irregular
  domains}.
\newblock \emph{\bibinfo{journal}{IEEE Signal Processing Magazine}}
  \textbf{\bibinfo{volume}{30}}, \bibinfo{pages}{83--98}
  (\bibinfo{year}{2013}).

\bibitem{ma2015}
\bibinfo{author}{Ma, J.}, \bibinfo{author}{Huang, W.},
  \bibinfo{author}{Segarra, S.} \& \bibinfo{author}{Ribeiro, A.}
\newblock \bibinfo{title}{{Diffusion filtering for graph signals and its use in
  recommendation systems}}.
\newblock In \emph{\bibinfo{booktitle}{Acoustics, Speech and Signal Processing
  (ICASSP), 2016 IEEE Int. Conf. on}}, \bibinfo{pages}{4563--4567}
  (\bibinfo{address}{Shanghai, China}, \bibinfo{year}{2016}).

\bibitem{segarra2015}
\bibinfo{author}{Segarra, S.}, \bibinfo{author}{Huang, W.} \&
  \bibinfo{author}{Ribeiro, A.}
\newblock \bibinfo{title}{Diffusion and superposition distances for signals
  supported on networks}.
\newblock \emph{\bibinfo{journal}{Signal Inform. Process. over Network., IEEE
  Trans. on}} \textbf{\bibinfo{volume}{1}}, \bibinfo{pages}{20--32}
  (\bibinfo{year}{2015}).

\bibitem{huang2015diffusion}
\bibinfo{author}{Huang, W.}, \bibinfo{author}{Segarra, S.} \&
  \bibinfo{author}{Ribeiro, A.}
\newblock \bibinfo{title}{Diffusion distance for signals supported on
  networks}.
\newblock In \emph{\bibinfo{booktitle}{Proc. Asilomar Conf. on Signals Systems
  Computers}}, \bibinfo{pages}{1219--1223} (\bibinfo{address}{Asilomar CA},
  \bibinfo{year}{2015}).

\bibitem{huang2016graph}
\bibinfo{author}{Huang, W.} \emph{et~al.}
\newblock \bibinfo{title}{Graph frequency analysis of brain signals}.
\newblock \emph{\bibinfo{journal}{J. Sel. Topics Signal Process.}}
  \textbf{\bibinfo{volume}{10}}, \bibinfo{pages}{1189--1203}
  (\bibinfo{year}{2016}).

\bibitem{spielman2009spectral}
\bibinfo{author}{Spielman, D.}
\newblock \bibinfo{title}{Spectral graph theory}.
\newblock \emph{\bibinfo{journal}{Lecture Notes, Yale University}}
  \bibinfo{pages}{740--0776} (\bibinfo{year}{2009}).

\bibitem{bassett2013robust}
\bibinfo{author}{Bassett, D.~S.} \emph{et~al.}
\newblock \bibinfo{title}{Robust detection of dynamic community structure in
  networks}.
\newblock \emph{\bibinfo{journal}{Chaos: An Interdisciplinary Journal of
  Nonlinear Science}} \textbf{\bibinfo{volume}{23}}, \bibinfo{pages}{013142}
  (\bibinfo{year}{2013}).

\bibitem{bassett2011dynamic}
\bibinfo{author}{Bassett, D.~S.} \emph{et~al.}
\newblock \bibinfo{title}{Dynamic reconfiguration of human brain networks
  during learning}.
\newblock \emph{\bibinfo{journal}{Proceedings of the National Academy of
  Sciences}} \textbf{\bibinfo{volume}{108}}, \bibinfo{pages}{7641--7646}
  (\bibinfo{year}{2011}).

\bibitem{Mucha2010}
\bibinfo{author}{Mucha, P.~J.}, \bibinfo{author}{Richardson, T.},
  \bibinfo{author}{Macon, K.}, \bibinfo{author}{Porter, M.~A.} \&
  \bibinfo{author}{Onnela, J.-P.}
\newblock \bibinfo{title}{Community structure in time-dependent, multiscale,
  and multiplex networks}.
\newblock \emph{\bibinfo{journal}{Science}} \textbf{\bibinfo{volume}{328}},
  \bibinfo{pages}{876--878} (\bibinfo{year}{2010}).

\bibitem{bassett2013multiscale}
\bibinfo{author}{Bassett, D.~S.} \& \bibinfo{author}{Siebenh{\"u}hner, F.}
\newblock \bibinfo{title}{Multiscale network organization in the human brain}.
\newblock \emph{\bibinfo{journal}{Multiscale Analysis and Nonlinear Dynamics:
  From Genes to the Brain}} \bibinfo{pages}{179--204} (\bibinfo{year}{2013}).

\bibitem{leonardi2013principal}
\bibinfo{author}{Leonardi, N.} \emph{et~al.}
\newblock \bibinfo{title}{Principal components of functional connectivity: a
  new approach to study dynamic brain connectivity during rest}.
\newblock \emph{\bibinfo{journal}{NeuroImage}} \textbf{\bibinfo{volume}{83}},
  \bibinfo{pages}{937--950} (\bibinfo{year}{2013}).

\bibitem{viviani2005functional}
\bibinfo{author}{Viviani, R.}, \bibinfo{author}{Gr{\"o}n, G.} \&
  \bibinfo{author}{Spitzer, M.}
\newblock \bibinfo{title}{Functional principal component analysis of fmri
  data}.
\newblock \emph{\bibinfo{journal}{Human brain mapping}}
  \textbf{\bibinfo{volume}{24}}, \bibinfo{pages}{109--129}
  (\bibinfo{year}{2005}).

\bibitem{bullmore2009complex}
\bibinfo{author}{Bullmore, E.} \& \bibinfo{author}{Sporns, O.}
\newblock \bibinfo{title}{Complex brain networks: graph theoretical analysis of
  structural and functional systems}.
\newblock \emph{\bibinfo{journal}{Nature Reviews Neuroscience}}
  \textbf{\bibinfo{volume}{10}}, \bibinfo{pages}{186--198}
  (\bibinfo{year}{2009}).

\end{thebibliography}

\newpage
\subsection*{Acknowledgements}
JDM acknowledges support from the Office of the Director at the National Institutes of Health through grant number 1-DP5-OD-021352-01. DSB acknowledges support from the John D. and Catherine T. MacArthur Foundation, the Alfred P. Sloan Foundation, the Army Research Laboratory and the Army Research Office through contract numbers W911NF-10-2-0022 and W911NF-14-1-0679, the National Institute of Health (2-R01-DC-009209-11, 1R01HD086888-01, R01-MH107235, R01-MH107703, R01MH109520, 1R01NS099348 and R21-M MH-106799), the Office of Naval Research, and the National Science Foundation (BCS-1441502, CAREER PHY-1554488, BCS-1631550, and CNS-1626008). The content is solely the responsibility of the authors and does not necessarily represent the official views of any of the funding agencies. 

\subsection*{Competing Interests}
The authors declare that they have no competing financial interests.

\subsection*{Correspondence}
Correspondence and requests for materials should be addressed to D.S.B.~(email: dsb@seas.upenn.edu).	

\end{document}